\documentclass[journal,12pt,onecolumn,draftclsnofoot,]{IEEEtran}

\newif\ifISIT

\ISITfalse 

\usepackage[utf8]{inputenc} 
\usepackage[T1]{fontenc}
\usepackage{url}              
\usepackage{cite}             

\usepackage[cmex10]{amsmath}  
\usepackage{mleftright}       
\mleftright                   

\usepackage{graphicx}         
\usepackage{booktabs}         

\usepackage{algorithm}
\usepackage{algpseudocode}

\usepackage{color}

\usepackage{amssymb, amsfonts}
\usepackage{bm}
\usepackage{comment}
\usepackage{enumerate}
\usepackage{subcaption}

\newtheorem{lemma}{Lemma}
\newtheorem{theorem}{Theorem}

\newtheorem{definition}{Definition}
\newtheorem{property}{Property}
\newtheorem{remark}{Remark}

\begin{document}

\ifISIT
\title{Computation of Marton's Error Exponent for Discrete Memoryless Sources} 
\else
\title{Computation of the optimal error exponent function for fixed-length lossy source coding in discrete memoryless sources}
\fi


%

\author{Yutaka Jitsumatsu,~\IEEEmembership{Member,~IEEE,}\thanks{Y.~Jitsumatsu is with Kyushu University.}
\thanks{This work was partially supported by JSPS KAKENHI Grant Numbers JP19K12156, JP23H00474, and JP23H01409.}
\thanks{A part of this paper was presented in ISIT2023.}
}

\maketitle

\begin{abstract}
Marton's optimal error exponent for the lossy source coding problem is defined as a non-convex optimization problem. This fact had prevented us to develop an efficient algorithm to compute it. This problem is caused by the fact that the rate-distortion function $R(\Delta|P)$ is potentially non-concave in the probability distribution $P$ for a fixed distortion level $\Delta$. The main contribution of this paper is the development of a parametric expression that is in perfect agreement with the inverse function of the Marton exponent. This representation has two layers. The inner layer is convex optimization and can be computed efficiently. 
The outer layer, on the other hand, is a non-convex optimization with respect to two parameters. We give a method for computing the Marton exponent based on this representation.
\end{abstract}

\section{Introduction}
Practical lossy source coding, such as audio coding, image coding, and video coding, achieves high compression ratios at the expense of quality degradation. The trade-off in the theoretical limit between compression rate and reproduced data quality is described by the rate-distortion function. 
Rate-distortion theory was originated by Shannon~\cite{Shannon1948, Shannon1959}.
See~\cite{Kieffer1993,Berger2003} for the development of rate-distortion theory.

The topic of this paper is a method for computing the error exponent in lossy source coding. The error exponent is a quantity that expresses the exponential decrease in the probability of the occurrence of distortion exceeding a pre-determined acceptable distortion level with the code length. 
The optimal error exponent is defined as a non-convex optimization problem.
This fact had prevented us to develop an efficient algorithm to compute it.

Consider a discrete memoryless source (DMS) with a probability distribution $P$ on a source alphabet $\mathcal{X}$. Let $\mathcal{Y}$ be a reproduced alphabet and assume that $\mathcal{X}$ and $\mathcal{Y}$ are finite sets. Assume that a symbol-wise distortion measure denoted by $d(x,y) \ge 0$ for $x\in \mathcal{X}$ and $y\in \mathcal{Y}$ is given\footnote{We assume that the distortion measure $d(x,y)$ satisfies 
    $\max_{x\in \mathcal{X}} \min_{y \in \mathcal{Y} } d(x,y) = 0$ and 
    $d_{\rm max} = \max_{(x,y)\in \mathcal{X\times Y} } d(x,y) <\infty$. This assumption is often made for technical reasons.
    \label{condition_distortion_measure}}. 
An $n$-symbol distortion is measured as $d(x^n, y^n) = (1/n) \sum_{i=1}^n d(x_i, y_i)$. 
Then, the rate-distortion function is given by 
\begin{align}
R(\Delta|P) = 
\min_{
\genfrac{}{}{0pt}{}{
V \in \mathcal{P(Y|X)}:
}{
\mathrm{E}_{P, V }[d(X,Y) ] \leq \Delta 
}
} I(P, V ), 
\label{Rate_distortion_function}
\end{align}
where $\mathcal{P(Y|X)}$ is the set of all conditional probability distribution on $\mathcal{Y}$ given $\mathcal{X}$, $I(P,V)$ is the mutual information, and $\Delta\ge 0$ is an acceptable distortion level. 
%
The optimal error exponent was established by Marton\cite{Marton1974}, which is expressed by
\begin{align}
    E_{\rm M}(R | \Delta, P) =
\min_{
\genfrac{}{}{0pt}{}
  {
    q_X \in \mathcal{P(X)} :
  }{
    R(\Delta | q_X) \geq R 
  }
}
D(q_X \| P),
\label{E_M}
\end{align}
for $0\leq R \leq R_{\max}(\Delta) := \max_{ q_X\in \mathcal{P(X)}} R(\Delta | q_X)$, where $D(q_X||P)$ is a relative entropy and
$\mathcal{P(X)}$ is the set of all probability distributions on $\mathcal{X}$.
Computation of Eq.~(\ref{E_M}) is the main topic of this paper.

\begin{figure}
\centering
\includegraphics[width = 0.4\textwidth]{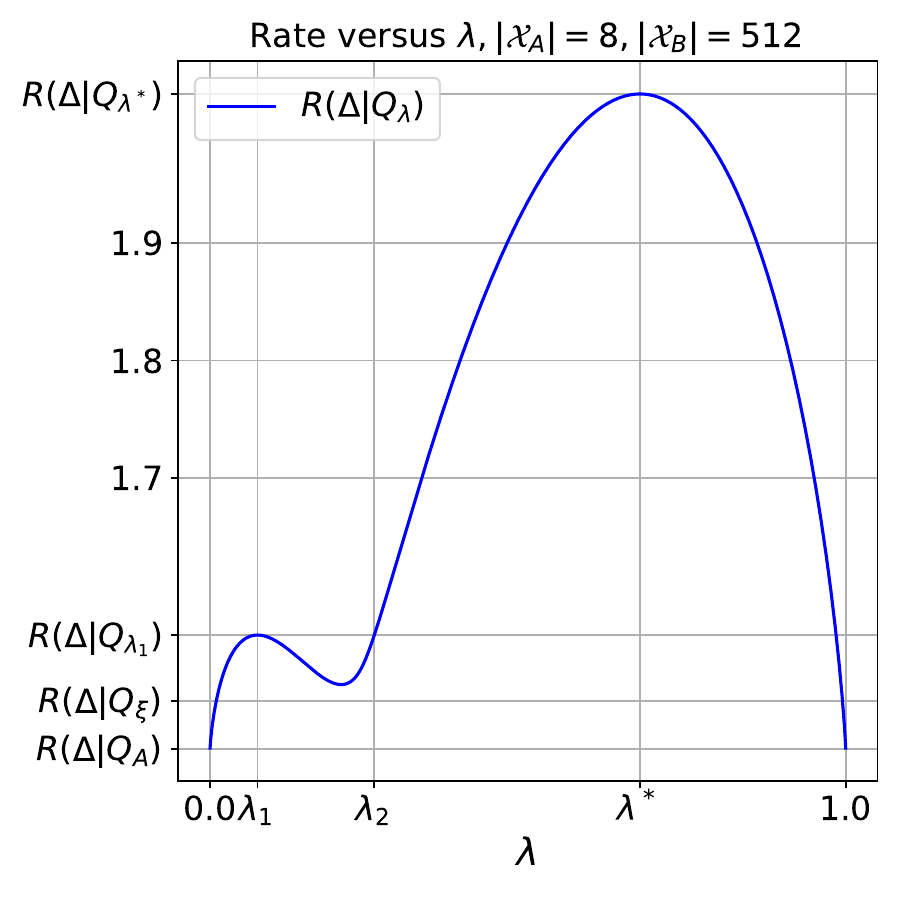}
\caption{The rate-distortion function $R(\Delta|Q_{\lambda})$ as a function of $\lambda$ of Ahlswede's counterexample with parameters \#1 in Table~\ref{table1}. The unit of rate $R$ is the bit.}
\label{fig.1}
\end{figure}

Until now, there was no algorithm to compute $E_{\rm M}(R | \Delta, P)$.
The main challenge in computing Marton's error exponent stems from the intrinsic nature of the rate-distortion function with respect to the probability distribution $P$.
In a typical example of $d(x,y)$, the rate-distortion function 
appears to be quasi-concave\footnote{
A function $f$ on $\mathcal{P(X)}$ is said to be quasi-convex if for all real numbers $\alpha \in \mathbb{R}$, the sublevel set $\{ P \in \mathcal{P(X)}: 
f(x) \le \alpha\}$ is convex. A function $f$ is quasi-concave if $-f$ is quasi-convex.
} with respect to $P$, just as the entropy function $H(P)$, the limit of lossless compression, is concave with respect to $P$.
In fact, in Marton's 1974 paper, it was not known whether the rate-distortion function was concave with respect to $P$ for any distortion measure $d(x,y)$. 
This problem was negatively solved by Ahlswede in 1990~\cite{Ahlswede1990}. 
The counter-example that he presented in his paper, even though he did not draw a figure, is shown in Fig.~\ref{fig.1}, where the source alphabet $\mathcal{X}$
is separated into two disjoint subsets $\mathcal{X}_A$ and $\mathcal{X}_B$ and 
$Q_\lambda \in \mathcal{P(X)}$ is a mixture of two uniform distributions
on $\mathcal{X}_A$ and $\mathcal{X}_B$ with mixture weights $\lambda $ and $1-\lambda$. 
The details will be explained in Section~\ref{sec:counterexample}. 
The function $R(\Delta| P )$ of this example clearly has two local maxima on the line $P=Q_\lambda$ for $0\le \lambda \le 1$.
Thus, we must assume that $R(\Delta|P)$ generally can have more than one local maxima with respect to $P$.
Notably, the distortion measure $d(x,y)$ in this counterexample is not an unrealistic extreme example, but a realistic one.

The optimization problem appeared in (\ref{E_M}) is classified as a non-convex programming because the objective function is a convex function, but the feasible set is not convex. A brute-force approach to finding (\ref{E_M}) is to sample a large number of $q_X \in \mathcal{P(X)}$,
computing $R(\Delta|q_X)$ to check if $R(\Delta|q_X)\geq R$, if so, computing $D(q_X\|P)$, 
and taking the maximum of $D(q_X\|P)$ to get an approximation for $E_{\rm M}(R | \Delta, P)$, 
which requires the computational cost to be exponential in $|\mathcal{X}|$.
There are no efficient general-purpose computational algorithms for non-convex programming problems, and general solution search methods such as the gradient method inevitably lead to local minima. 
Therefore, in many cases, the only solution is to start the search from several initial values and select the smallest one among the local minima obtained. This is true for general non-convex programming problems, but what about the specific case of Marton's exponent? 
It would be great if we could avoid falling into local minima and calculate the Marton exponent function correctly in an efficient way.

The contribution of this paper is to give a theoretically guaranteed method for computing the Marton exponent. To this aim, we develop a parametric expression for the inverse function of the Marton exponent, denoted by $R_{\rm M}(E|\Delta, P)$.
Our proposed method is a two-layer computation method based on this parametric expression. The inner optimization problem is a convex optimization problem with respect to the probability distribution on $\mathcal{Y}$ and can be solved by an efficient algorithm. The outer optimization problem with respect to two parameters $(\nu, \mu)$ is solved by the brute-force method.
Some may think that the brute-force method is not an efficient computational method. However, considering that the original problem is a non-convex programming problem and that the brute-force method requires a computational complexity on the order of the exponent of the alphabet size $|\mathcal{X}|$, nonlinear optimization in only two variables is the minimum cost to pay. 
It is a small price to pay for guaranteeing optimality.

\subsection{Sub-optimality of Blahut's exponent}

Readers familiar with the Arimoto-Blahut algorithm~\cite{Arimoto1972, Blahut1972} and its extensions~\cite{DupuisISIT2004,Cheng2005,YasuiISIT2007} may point out that Arimoto~\cite{Arimoto1976} gave an algorithm for the exponent. However, a careful reading of Arimoto's paper~\cite{Arimoto1976} reveals that his algorithm for computing the error exponent of lossy source coding of a DMS is based on Blahut's exponent $E_{\rm B}(R|\Delta,P)$~\cite{Blahut1974}, defined by the following formula. 
\begin{align}
E_{\rm B}(R | \Delta, P)
=
\sup_{\rho\geq 0} 
\left[ \rho R +
\inf_{\nu\geq 0} 
\left\{
\rho \nu \Delta 
-
\min_{p_Y} 
\log \sum_{x \in \mathcal{P(X)} }
P(x) 
\left(
\sum_{y\in \mathcal{Y}} p_Y(y) {\rm e}^{-\nu d(x,y) }
\right)^{-\rho}
\right\}
\right] \label{def:E_B}
\end{align}
for $0\leq \Delta \leq \Delta_{\rm max}$ and $0\leq R\leq R_{\max}(\Delta)$. 

The relationship between Marton's exponent and Blahut's exponent 
can be stated as follows.
\begin{lemma}\label{lemma.1}
For any $P\in \mathcal{P(X)}$, distortion measure $d(x,y)$, $R\geq 0$, and $\Delta\geq0$,
$E_{\rm B}(R|\Delta,P)$ 
is a lower convex envelope of $E_{\rm M}(R|\Delta,P)$.
\end{lemma}
The proof of Lemma~\ref{lemma.1} is given in Section~\ref{Section:Proof}.

\begin{figure}
\centering
\includegraphics[width=0.4\textwidth]{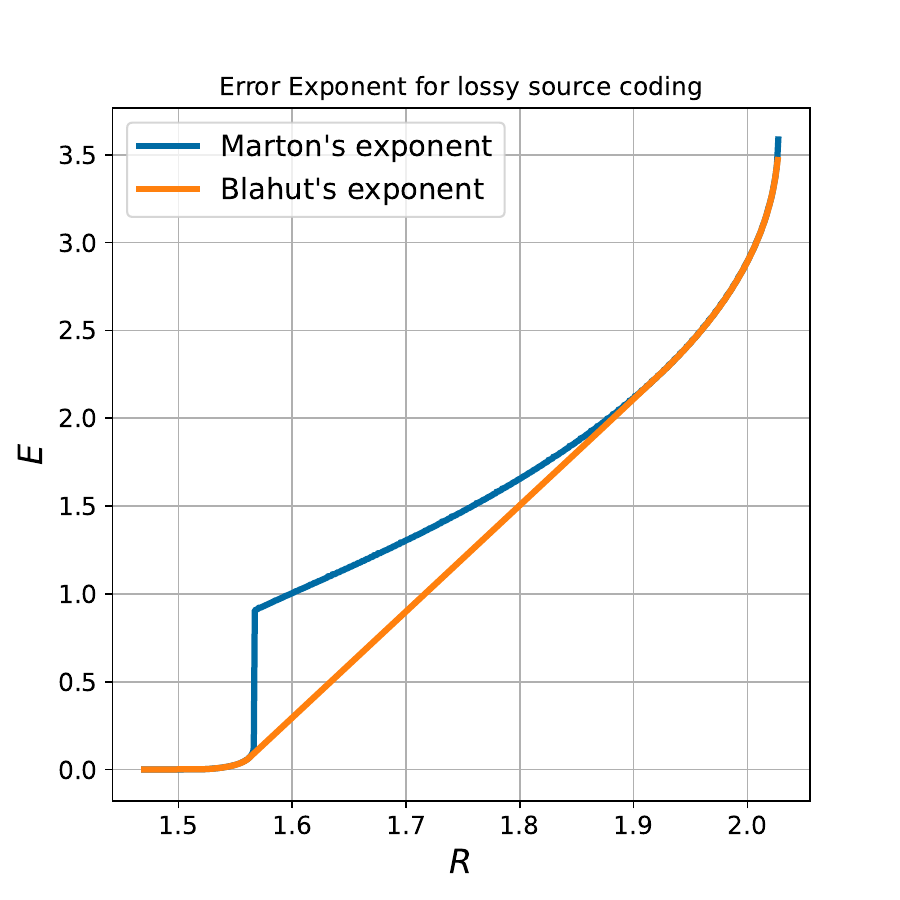}
\caption{Marton's and Blahut's error exponents are illustrated 
as functions of $R$ for Ahlswede's counterexample with parameters \#1 in Table~\ref{table1}.  
}
\label{fig.error_exponent}
\end{figure}

Let us look at a numerical example. 
Fig.~\ref{fig.error_exponent} shows the graphs of two error exponents for the example given by Ahlswede in Fig.~\ref{fig.1}. 
Blahut's exponent is a continuous function of $R$.
On the other hand, Marton's exponent jumps at some $R$.
How to draw these graphs is explained in Section~\ref{sec:counterexample}.
We observe that the jump occurs when the feasible region $ \{q_X \in \mathcal{P(X)} : R(\Delta | q_X ) \ge R\}$  changes discontinuously for small changes in $R$.
In summary, the main reason why there is no efficient way to compute the Marton exponent is that the rate-distortion function is generally not concave with respect to $P$.

\subsection{Paper organization}
The rest of this paper is organized as follows.
Section II presents the main results, which consist of four subsections. Theorem 1 in Section II-A gives a parametric expression for the inverse of the Marton exponent. An algorithm for computing the inverse of the Marton exponent based on Theorem 1 is provided in Section II-B. 
In Sections II-C and II-D, we analyze the computational cost of the proposed method and evaluate it in comparison with the computation of the Blahut exponent.
In Section III, a numerical example demonstrates the effectiveness of the proposed computational method and shows the potential non-convexity of the rate-distortion function. Section IV gives the proof of Theorem 1. 
In this paper, $\log$ denotes the natural logarithm unless otherwise noted.

\section{Main Results}
\subsection{Parametric expression of the inverse function of the Marton exponent}
According to the standard approach to solving the optimization problem with equality and inequality constraints, we can introduce the Lagrange multipliers for the constraints and consider an unconstrained optimization.
The Lagrange dual of the primal problem of (\ref{E_M}) is given by\footnote{The expression (\ref{eq.6}) appeared in~\cite[Eq.~(30) of Theorem 2]{ArikanMerhav1998} and it was proved that this function is the lower convex envelope of $E_{\rm M}(R|\Delta, P)$. It is obvious from Lemma~\ref{lemma.1} that (\ref{eq.6}) coincides with (\ref{def:E_B}).}
\begin{align}
    \sup_{\rho \ge 0} \min_{q_X\in \mathcal{P(X)}} \{ D(q_X\|P) + \rho [ R - R(\Delta|q_X) ] \}. 
    \label{eq.6}
\end{align}
However, there may be a gap between the primal problem (\ref{E_M}) and (\ref{eq.6}) since $R(\Delta|q_X)$ is not necessarily concave in $q_X$. 

How can we solve this seemingly hopeless computation of the Marton exponent? 
The first approach is simple: we consider the inverse function of the Marton exponent:
\begin{align}
R_{\rm M}(E | \Delta, P )
=
\max_{
\genfrac{}{}{0pt}{}{
q_X \in \mathcal{P(X)}:
}{
D(q_X\|P) \leq E
}
} R(\Delta | q_X).  \label{R_M}
\end{align}
Now the feasible region is convex and the objective function is non-concave. 
Several steps are still needed to successfully handle the non-concavity of $R( \Delta|q_X)$ in $q_X$. After going through those steps, we obtain the next theorem. 

\begin{theorem}
\label{theorem2}
For any $P\in \mathcal{P(X)}$, $E\ge 0$, and $0\leq \Delta \leq \Delta_{\max}
:={\min_y} \sum_{x\in \mathcal{X}} P(x) d(x,y)$, we have
\begin{align}
R_{\rm M}(E| \Delta, P ) =
\sup_{\nu\geq 0} 
\inf_{\mu \geq 0}
\left[
- \nu \Delta + \mu E 
+ \min_{p_Y \in \mathcal{P(Y)}} G^{(\mu, \nu)}(p_Y|P)
\right]. \label{eq.theorem2}
\end{align}
where 
\begin{align}
G^{(\mu,\nu)} ( p_Y | P) =
\begin{cases}
\mu \log \sum_x P(x) \left\{ \sum_{y} p_Y (y) \mathrm{e}^{-\nu d(x,y)} \right\}^{-1/\mu} & \text{ if } \mu > 0,\\
- \log \min_x \sum_y p_Y(y) \mathrm{e}^{-\nu d(x,y)} &\text{ if } \mu = 0.
\end{cases} 
\label{Gmunup_YP2}
\end{align}
\end{theorem}
The proof of Theorem~\ref{theorem2} is given in Section~\ref{section:proof:theorem2}.

The expression for $R_{\rm M}(E|\Delta, P)$ is worth noting that the function $G^{(\mu, \nu)}(p_Y|P)$ in (\ref{eq.theorem2}) is convex in $p_Y$.
This property directly implies an efficient computational algorithm.
Also, (\ref{eq.theorem2}) is an expression that is perfectly consistent with the inverse function of Marton's optimal error exponent. 
To the best of our knowledge, such an exact parametric expression has not been known before.


The function $R_{\rm M}(E | \Delta , P )$ satisfies the following properties.
In particular, the continuity property is important.
\begin{property} ~\label{property_R_M}
\begin{enumerate}[a)]
\item $R_{\rm M}(0 | \Delta , P ) = R(\Delta|P)$ holds. 
\item $R_{\rm M}(E | \Delta , P )$ is a monotonically non-decreasing function of $E\ge 0$ for fixed $\Delta\geq 0$
and $P$.
\item $R_{\rm M}(E | \Delta , P ) = \max_{q_X\in \mathcal{P(X)} } R(\Delta|q_X)$ for $E \geq  E_{\rm max} :=
\min_{ q_X \in \arg \max_{ q'_X} R(\Delta| q'_X)   } D(q_X||P) $. 
\item $R_{\rm M}(E | \Delta , P )$ is a continuous function of $E\ge 0$ for fixed $\Delta\geq 0$
\end{enumerate}
\end{property}

\textit{Proof}: 
Property~\ref{property_R_M} a) holds since $D(q_X\|P)=0$ implies $q_X=P$.
For Property~\ref{property_R_M} b), the monotonicity is obvious from the definition. 
Property~\ref{property_R_M} c) holds because if $E\ge E_{\rm max}$, we can choose 
an optimal $q_X$ that achieves the global maximum of $R(\Delta|q_X)$. 
See Appendix~\ref{sec:appendix:continuity} for the proof of Property~\ref{property_R_M} d).

The idea of analyzing the inverse function of the error exponent was first introduced by Haroutunian et al.~\cite{Haroutunian1984,Harutyunyan2004}. They defined the rate-reliability-distortion function as the minimum rate at which the encoded messages can be reconstructed by the decoder with exponentially decreasing probabilities of error. They proved that the optimal rate-reliability-distortion function is given by (\ref{R_M}).

\subsection{Computation Method for $R_{\rm M}(E|\Delta, P)$}
Our proposed method for computing $R_{\rm M}(E|\Delta, P)$ based on Theorem~\ref{theorem2} is as follows. 
We discretize the search range for $(\mu,\nu) \in [0,+\infty]^2$ into an appropriate grid. Let $\Delta\mu$ and $\Delta\nu$ be the tick widths and the number of grids be $M$ and $N$, respectively.
Let $\mu_i = i (\Delta \mu)$ and $\nu_j = j (\Delta \nu)$ for $i \in\{0,1,\ldots, M-1\}$
and $j \in\{0,1,\ldots, N-1\}$ and we compute 
\begin{align}
  \max_{0\le j \le N-1}
  \min_{0\le i \le M-1}
  \left[
  - \nu_j \Delta
  + \mu_i E + \min_{p_Y} G^{(\mu_i, \nu_j)} (p_Y| P)
  \right] \label{approximateR_M_E}
\end{align}
as an approximation of (\ref{eq.theorem2}). 
Minimization with respect to $p_Y$ in the block bracket in (\ref{approximateR_M_E}) is a convex programming problem if $\mu_i>0$ and a linear programming problem if $\mu_i=0$, so it can be obtained according to conventional methods.
We use Arimoto's algorithm~\cite{Arimoto1976} with $\rho=1/\mu_i$ if $\mu_i>0$. 
If $\mu_i=0$, minimization of $G^{(\mu_i,\nu_j)}(p_Y|P)$ reduces to the following linear programming problem: 
\begin{align}
\text{maximize} \quad  & c \label{objective_function}\\
\text{subject to} \quad &\sum_y p_Y(y) \mathrm{e}^{-\nu_j d(x,y)} \ge c, \quad ^\forall x\in\mathcal{X}, \label{constraint1}\\
& p_Y(y)\geq 0,\label{constraint2}\\
& \sum_{y\in \mathcal{P(Y)}} p_Y(y) = 1, \label{constraint3}
\end{align}
where variables are $p_Y(y)$ and $c$. 
Since linear programming problems are not specific to information theory, a general-purpose solver is used. 
A detailed description is given in Algorithm~\ref{proposed_algorithm}.
To make the paper self-contained, the Arimoto algorithm is described in Algorithm~\ref{algorithm1}.

\begin{algorithm}[t]
  \caption{Proposed algorithm for computing the inverse function of the optimal error exponent in lossy source coding}
  \label{proposed_algorithm}
  \begin{algorithmic}[1]
  \Procedure{Computation\_Inverse\_of\_Error\_Exponent }{$P, d, \Delta$}
    \State{Set $M, N, K$ and $\varDelta \mu, \varDelta \nu, \varDelta E$ according to the precision.}
        \For{$j=0$ to $N-1$} 
            \State{$\nu_j \gets j(\varDelta \nu)$}
            \State{Solve (\ref{objective_function}) to (\ref{constraint3})
            to obtain the optimal value $c^*$ and $G^{(0,\nu_j)}(P) \gets -\log c^*$.}
            \For{$i=1$ to $M-1$}
                \State{$\mu_i \gets i (\varDelta \mu)$. } 
                \State{Execute Algorithm~\ref{algorithm1} with $\rho = 1/\mu_i$, $\nu = \nu_j$ to obtain an optimal }
                \State{distribution $p_Y^*$. Substitute $p_Y^*$ into (\ref{Gmunup_YP2}) to obtain $g_{i,j} \gets G^{(\mu_i, \nu_j)}(p_Y^*|P) $ }
            \EndFor
        \EndFor
    \For{$k=0$ to $K-1$}
        \State{$E_k = k (\varDelta E)$}
        \For{$j=0$ to $M-1$} 
            \State{$a_{j,k} \gets \displaystyle \min_{i\in\{0, \ldots, N-1\} } \{ g_{i,j} + \mu_i E_k\} $}
        \EndFor
        \State{$R_{\rm M}(E_k| \Delta, P) \gets \displaystyle \max_{j\in \{0, \ldots, M-1\} } \{ a_{j,k} - \nu_j \Delta \}$ }
    \EndFor
  \EndProcedure
  \end{algorithmic}
\end{algorithm}

\begin{algorithm}[t]
  \caption{Arimoto algorithm for computing the error exponent of lossy source coding~\cite{Arimoto1976}}
  \label{algorithm1}
  \begin{algorithmic}[1]
  \Procedure{Arimoto\_Lossy\_Source\_Coding\_Error\_Exponent}{$\rho, \nu, P, d$}
  \State{$ \epsilon>0$, and ${\rm max\_itr} >0$ are given.}
  \State{$ i \gets 0$}
  \State{Choose initial output distribution $p_Y^{[0]}$ arbitrarily so that all elements are nonzero.}
  \Repeat
  \begin{align}
  &\hspace{-6mm} q_{Y|X}^{[i]}(y|x) = \frac{ p_Y^{[i]}(y) \mathrm{e}^{-\nu d(x,y)} }{ \sum_{y\in \mathcal{Y} } p_Y^{[i]}(y) \mathrm{e}^{-\nu d(x,y)}} \\
  &\hspace{-6mm}p_Y^{[i+1]}(y) = 
  \frac{ \Big[ {\displaystyle \sum_x } P(x) {\rm e}^{\rho\nu d(x,y) } q_{Y|X}^{[i]}(y|x)^{1+\rho} 
  \Big]^{\frac{1}{1+\rho} } }
  { { \displaystyle \sum_{y'}}
  \Big[ {\displaystyle \sum_x } P(x) {\rm e}^{\rho\nu d(x,y') } q_{Y|X}^{[i]}(y'|x)^{1+\rho} 
  \Big]^{\frac{1}{1+\rho} } 
  }\\
  &\hspace{-6mm} i \gets i+1
  \end{align}
  \Until{$\left\lvert 
 G^{(1/\rho, \nu)} (p_Y^{[i]} | P )
 -
 G^{(1/\rho, \nu)} (p_Y^{[i-1]} | P )
  \right\rvert < \epsilon$ or $i > {\rm max\_itr} $}
  \EndProcedure
  \end{algorithmic}
\end{algorithm}

\ifISIT
The correct approach to the optimization problem is to find a solution that satisfies the 
Karush–Kuhn–Tucker (KKT) condition and consider the Lagrangian function. 
To do this, we need to evaluate the derivative of $R(\Delta|q_X)$ w.r.t. $q_X$.
Because $R(\Delta|q_X)$ is defined by a constrained optimization problem (\ref{Rate_distortion_function}), another Lagrangian is introduced. 
The author 
was unable to derive a parametric formula that is in exact agreement with Marton's formula. 
We will take a different approach to compute Marton's exponent in Section III.
Since $R(\Delta|q_X)$ determined by the solution of the optimization problem (\ref{Rate_distortion_function}) is used in the constraint, we find that the result of directly substituting (\ref{Rate_distortion_function}) into (\ref{E_M}) is very complicated. 
\else

\fi

\subsection{Comparison with the inverse function of the Blahut exponent}
To illustrate the difference between the Marton and Blahut exponents, the inverse function of the Blahut exponent is introduced here. 
The Blahut exponent (\ref{def:E_B}) is convex and monotone non-decreasing in $R$. 
The parameter $\rho$ in (\ref{def:E_B}) is recognized as the slope of a supporting line to the curve $E_{\rm B}(R|\Delta, P)$. Let $\mu = 1/\rho$ be the slope of 
the inverse of the Blahut exponent and we allow $\mu$ to take zero.
Then, the inverse of $E_{\rm B}(R|\Delta, P)$
can be expressed by the following formula. 
\begin{theorem} \label{theorem_inverse_Blahut}
The inverse of the Blahut exponent is expressed as 
\begin{align}
    R_{\rm B}(E|\Delta, P)
=\inf_{\mu\ge 0} \sup_{\nu\ge 0}
\left[ 
\mu E - \nu \Delta  + \min_{p_Y \in \mathcal{P(Y)} }
G^{(\mu, \nu)}(p_Y|P) \right].
\label{def:R_B}
\end{align}
\end{theorem}

See Appendix~\ref{appendix_proof_theorem_inverseBlahut} for the proof. 

The difference between the parametric expressions (\ref{eq.theorem2}) and (\ref{def:R_B}) for the inverse functions of the Marton and the Blahut exponents is the order in which they take the infimum with respect to $\mu$ and taking the supremum with respect to $\nu$. 
In Section~\ref{section:numerical_examples}, we will see that this difference leads to a critical gap between the two exponents.




\subsection{Computational cost}
\label{remark3}

The proposed computation method consists of outer exhaustive search over $(\mu, \nu)$ and inner convex optimization. 
The outer exhaustive search to construct an array of size $N\times M$ requires $N M$ evaluations of Arimoto algorithm, as shown in 3 to 11 lines of Algorithm~\ref{proposed_algorithm}. For the inner optimization, we use the conventional Arimoto algorithm~\cite{Arimoto1976}.  

The approximation error at the $t$-th iteration of the Arimoto algorithm is upper bounded by $\frac{(1+\rho)}{t} \log |\mathcal{Y}| $~\cite{Arimoto1976}. Therefore, for a fixed tolerance $\epsilon>0$, the number of iterations is $O(\frac{(1+\rho) \log |\mathcal{Y}| }{\epsilon})$. The computational cost for one iteration is $O( | \mathcal{X} | | \mathcal{Y} | )$. Therefore, the computational cost of running the Arimoto algorithm once is $O( |\mathcal{X}| |\mathcal{Y}| \frac{\log |\mathcal{Y}|}{ \epsilon} )$. Then the total cost for lines 3 through 11 is $O( N M |\mathcal{X}| |\mathcal{Y}| \frac{\log |\mathcal{Y}|}{ \epsilon} )$. 
The computational cost for lines 12 through 18 of the algorithm~\ref{proposed_algorithm} is $O(NMK)$.
Finally, we evaluate that the computational cost for Algorithm~\ref{algorithm1} is $O(NM ( K + |\mathcal{X}| |\mathcal{Y}| \frac{\log |\mathcal{Y}|}{ \epsilon}))$.
In order to know the required values of $N$ and $M$ values, the approximation error (\ref{approximateR_M_E}) must be evaluated, but this is left as a future work.

Would the computational cost of finding the Blahut index (\ref{def:E_B}) or its inverse function (\ref{def:R_B}) be less than this?
Arimoto's algorithm (Algorithm~\ref{algorithm1}) efficiently finds the optimization for $p_Y$ with fixed $\rho$ and $\nu$ in the parametric representation (\ref{def:E_B}), but to find the Blahat index, the optimization for $\rho$ and $\nu$ must be performed. 
This situation is the same for the Marton exponent and the Blahut exponent. 
To the best of the authors' knowledge, there is no efficient way to solve the outer optimization for Blahut's exponent\footnote{Hayashi's recent paper on the EM algorithm based on Bregman divergence~\cite{Hayashi_Bregman_Divergence} is noteworthy because one application is to obtain the rate-distortion function, which, unlike the Arimoto-Blahut algorithm, does not require optimization with respect to Lagrange multipliers. We expect future research in this direction.}, and a naive grid-based method will incur a computational cost that is the same as that of the proposed method.

\section{Numerical Examples}
\label{section:numerical_examples}
In this section, the validity of the calculation method proposed in the previous section is verified by numerical examples.
The rate-distortion function of an example created without careful consideration is in many cases a quasi-concave function with respect to the probability distribution. This cannot demonstrate the effectiveness of the proposed method, since $E_{\rm M}(R|\Delta, P)$ coincides with $E_{\rm B}(R | \Delta, P)$. Therefore, we will use the counterexample discovered by Ahlswede~\cite{Ahlswede1990}.
Ahlswede gave an example where the rate-distortion function has two local maxima, but he did not provide specific values or graphs in his paper. In this paper, we have seen a numerical example of the rate-distortion function in Fig.~\ref{fig.1} of Introduction.
The next subsection provides the exact definition of the source and distortion measure for this numerical examples.
The graph has the effect of making it easier to grasp visually and intuitively. In the special case of Ahlswede's counterexample, we show that the Marton exponent can be calculated directly by applying the formula (\ref{E_M}).

\label{sec:counterexample}
%

Ahlswede's counterexample is defined as follows:
Let $\mathcal{Y}=\mathcal{X}$ and $\mathcal{X}$ is partitioned into $\mathcal{X}_A$ and $ \mathcal{X}_B$. 
Define the distortion measure as
\begin{align}
d(x,y) 
= 
\begin{cases}
  0, \text{ if } x=y \in \mathcal{X}, \\
  1, \text{ if } x\neq y \text{ and } x, y \in \mathcal{X}_A,\\
  a, \text{ if } x\neq y \text{ and } x, y \in \mathcal{X}_B,\\
  b, \text{ otherwise. } 
\end{cases}
\label{distortion}
\end{align}
The constant $b$ is a sufficiently large value so that encoding a source output $x\in \mathcal{X}_A$ into $y \in \mathcal{X}_B$ or vice versa has a large penalty. 
The constant $a$ is determined later.
It can be seen that the distortion measure (\ref{distortion}) is not an odd situation and can be adapted to situations where it is necessary to distinguish almost perfectly whether $x$ is in $\mathcal{X}_A$ or $\mathcal{X}_B$.

Assume $|\mathcal{X}_B| = | \mathcal{X}_A|^3$, 
where $|\cdot |$ denotes the cardinality of a set.
Let $Q_A$ and $Q_B$ be uniform distributions on $\mathcal{X}_A$ and $\mathcal{X}_B$, that is, 
\begin{align}
Q_A(x) &= \begin{cases}
1/|\mathcal{X}_A|, &\text{ if } x \in \mathcal{X}_A, \\
0, & \text{ if } x \in \mathcal{X}_B,
\end{cases} \\
Q_B(x) &= \begin{cases}
0, &\text{ if } x \in \mathcal{X}_A, \\
1/|\mathcal{X}_B|, & \text{ if } x \in \mathcal{X}_B. 
\end{cases}
\end{align}
For $\lambda\in [0,1]$, we denote $Q_\lambda = \lambda Q_A+(1-\lambda)Q_B$.
The rate-distortion function of $Q_A$ and $Q_B$ are 
\begin{align}
R(\Delta | Q_A) &= \log | \mathcal{X}_A | - h(\Delta) - \Delta \log ( | \mathcal{X}_A | - 1 ),\\ 
R(\Delta | Q_B) &= \log | \mathcal{X}_B | - h( \textstyle \frac{\Delta}{a}  ) - \frac{\Delta}{a}  \log ( | \mathcal{X}_B | - 1 ). 
\end{align}
To simplify the calculation, Ahlswede chose 
the parameters $a$ and $\Delta$ so that 
\begin{align}
\textstyle \frac{\Delta}{a} & =  1 - \Delta, \label{def_a} 
\end{align}
\begin{align}
& \log | \mathcal{X}_A | - \Delta \log ( | \mathcal{X}_A | - 1 )\notag\\
& =
\log | \mathcal{X}_B | - (1-\Delta)  \log ( | \mathcal{X}_B | - 1 )
\label{def_Delta}
\end{align}
hold.

The conjecture that $R(\Delta|P)$ is quasi-convex in $P$ for any given $d(x,y)$ and $\Delta$ is disproved if $R(\Delta|P)$ is not quasi-convex on any convex subset of $\mathcal{P(X)}$ for some $d(x,y)$ and some $\Delta$.  
Using the distortion function (\ref{distortion}) and the parameters $a,\Delta$
determined by (\ref{def_a}), (\ref{def_Delta}), 
Ahlswede analyzed the rate-distortion function $R(\Delta|P)$ for 
$P \in \{ Q_\lambda= \lambda Q_A + (1-\lambda) Q_B: 0\leq \lambda\leq 1\} \subset \mathcal{P(X)}$
and showed that 
if $|\mathcal{X}_A|$ is sufficiently large, 
$R(\Delta | Q_{\lambda} ) $ has local maximum different from the global maximum. 
This suggests that $R(\Delta | P ) $ of this case is not quasi-concave in $P$. 

\begin{table}
    \caption{Parameters for numerical evaluation}
    \centering
    \begin{tabular}{ccc} \hline 
         & \#1 & \#2 \\\hline 
    $|\mathcal{X}_A|$  &  $8$ & $50$ \\\hline 
    $|\mathcal{X}_B|$ & $512$ & $2500$\\\hline 
    $\Delta$     & $0.254$ & $0.333$ \\\hline 
    $a$ & $0.340$ & $0.501$\\\hline 
    $P$ & $0.01 Q_A+0.99 Q_B$ & $0.2 Q_A + 0.8Q_B$\\\hline 
    \end{tabular}
    \label{table1}
\end{table}

Numerical simulations were performed using the parameters listed in Table 1. Initially, the first parameters labeled "\#1" were used.
Fig.~\ref{fig.1} in Introduction is the function $R(\Delta | Q_{\lambda} ) $ for $\lambda \in \{0, 0.0001, 0.0002, \ldots, 0.9999, 1\}$, each of which is computed by the Arimoto-Blahut algorithm. 
The unit of rate $R$ is the bit in all graphs in this section.
If $|\mathcal{X}|$ is smaller than $8$, the graph of $R(\Delta|Q_{\lambda})$ does not have a local maximum that is different from the global maximum. 
We observe $R(\Delta |Q_\lambda)$ is bimodal with global maximum at $\lambda = \lambda^* = 0.676$
and local maximum at $\lambda = \lambda_1 = 0.0746$.


Next, let us draw the graph of the error exponent using the rate-distortion function in Fig.~\ref{fig.1}.
We give the following theorem to evaluate the error exponent for Ahlswede's counterexample.
\begin{theorem}
\label{theorem1}
Assume the distortion measure $d(x,y)$ is given by (\ref{distortion}) and
let $P=Q_\xi$ for a fixed $\xi\in [0,1]$.
Then, we have
\begin{align}
E_{\rm M}(R | \Delta, Q_{\xi})
=
\min_{
\genfrac{}{}{0pt}{}{
\lambda \in[0,1]:
}{
R(\Delta |Q_\lambda) \geq R
}
}
D_2(\lambda \| \xi )
\end{align}
where $D_2(p\|q) = p\log\frac{p}{q} + (1-p) \log \frac{1-p}{1-q}$ is a binary divergence.
\end{theorem}

See Appendix A for the proof.

Theorem~\ref{theorem1} ensures that the optimal error exponent can be computed as follows:

{\bf [Computation method of the error exponent for Ahlswede's counterexample]}

Let $N$ be a large positive integer and let $\lambda_i = i/N$ for $i=0,1,\ldots, N$.
Compute $R_i = R(\Delta|Q_{\lambda_i})$ and $D_i = D_2(\lambda_i \| \xi )$.
Then, arrange $(R_i, D_i)$ in ascending order of $R_i$.
Put $E_i = \min_{j\geq i} D_j$.
Then, by plotting $(R_i, E_i)$ for $i=0,1,\ldots, N$, we obtain
the graph of $E=E_{\rm M}(R| \Delta, Q_{\xi})$ for $R(\Delta|Q_\xi)\leq R \leq R_{\max}$.
We can add a straight line segment $E=0$ for $0\leq R \leq R(\Delta|Q_\xi)$.


Fig.~\ref{fig.error_exponent} shows the error exponent for Ahlswede's counterexample of Fig.~\ref{fig.1}.
The probability distribution of the source is chosen as 
$P=Q_{\xi}$ with $\xi = 0.01$. 
We observe that $E_{\rm M}(R|\Delta, P) =0$ for $R\leq R(\Delta|Q_{0.01}) = 1.510$
and $E_{\rm M}(R|\Delta, P)$ gradually increases for $1.510\leq R \leq R(\Delta|Q_{ \lambda_1})=1.566$. 
At $R=1.566$, the curve jumps from $E = D(Q_{\lambda_1} \| Q_{\xi}) =0.126$
to $E = D( Q_{\lambda_2} \| Q_\xi) = 0.904$, where 
$\lambda_2 = 0.258$ satisfies $ R(\Delta|Q_{\lambda_1}) = R(\Delta | Q_{\lambda_2})$. 
For $ R(\Delta|Q_{\lambda_1}) < R \leq R(\Delta |Q_{\lambda^*})$,
the graph is expressed by $(R,E) = ( R(\Delta|Q_\lambda), D(Q_\lambda\|Q_\xi) )$ with $ \lambda \in (\lambda_2, \lambda^*)$. 

In Fig.~\ref{fig.error_exponent}, Blahut's parametric expression (\ref{def:E_B}) 
of error exponent is also plotted, where optimal distribution $p_Y^*$
for (\ref{def:E_B}) is computed by Algorithm~\ref{algorithm1}.
This figure clearly shows that there is a gap between these two exponents.

\begin{figure}
    \centering
    \includegraphics[width=0.4\textwidth]{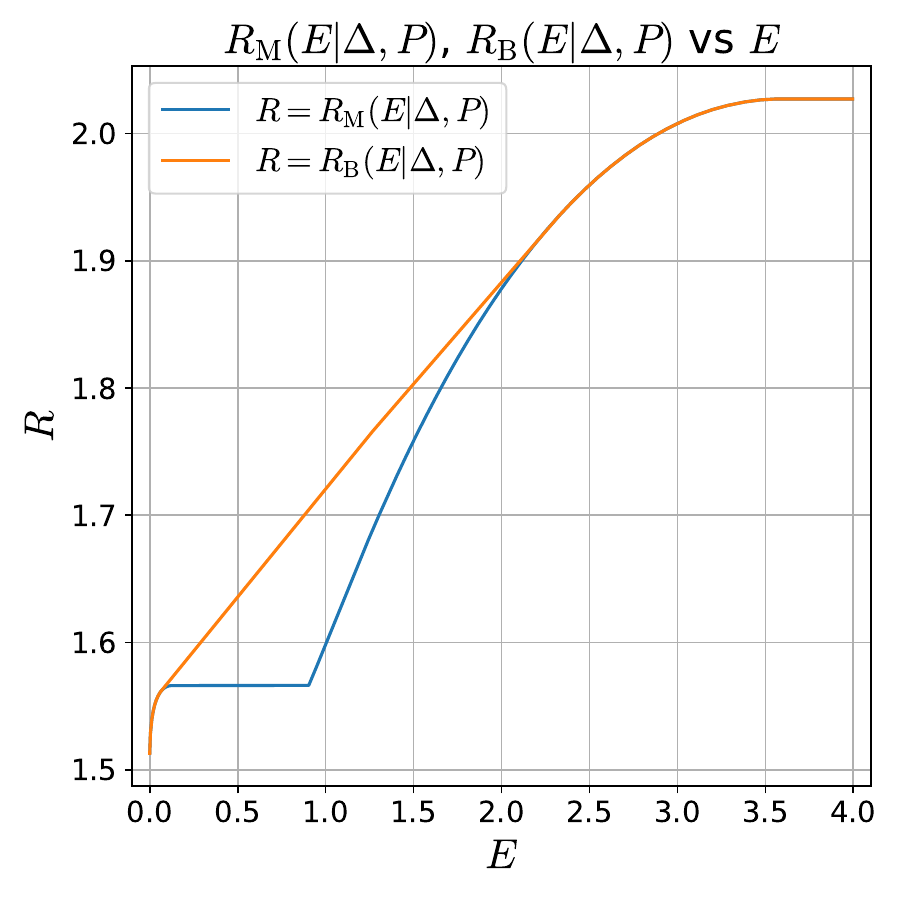}
    \caption{$R_{\rm M}(E | \Delta , P)$ for Ahlswede's counterexample of Fig.~\ref{fig.1}.
    }
    \label{fig.inverse}
\end{figure}

The computation results of the proposed method is shown in Fig.~\ref{fig.inverse}, where the parameters are the same as
Fig.~\ref{fig.error_exponent}. We confirm that $R_{\rm M}(E|\Delta, P)$ is correctly computed. 
Note that while Marton's exponent in Fig~\ref{fig.error_exponent} was computed based on Theorem~\ref{theorem1}, which holds only for Ahlswede's counterexamples, the proposed method is applicable to any $P$, $d$, and $\Delta$.
In Fig.~\ref{fig.inverse}, the inverse of Blahut's exponent is also shown. 

\begin{figure}
    \begin{minipage}[b]{0.5\linewidth}
        \centering
        \includegraphics[width=0.99\linewidth]{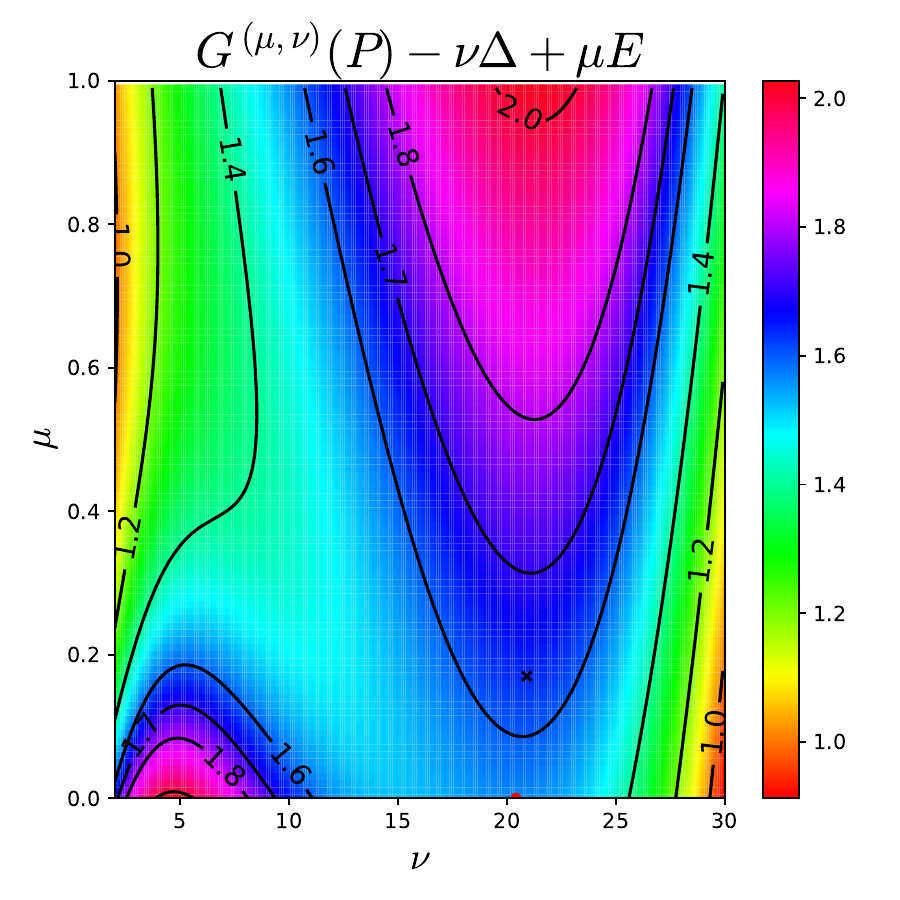}
        \subcaption{$E=0.5$}\label{fig:non_convexity_mu_nu_left}
    \end{minipage}
    \begin{minipage}[b]{0.5\linewidth}
        \centering
        \includegraphics[width=0.99\linewidth]{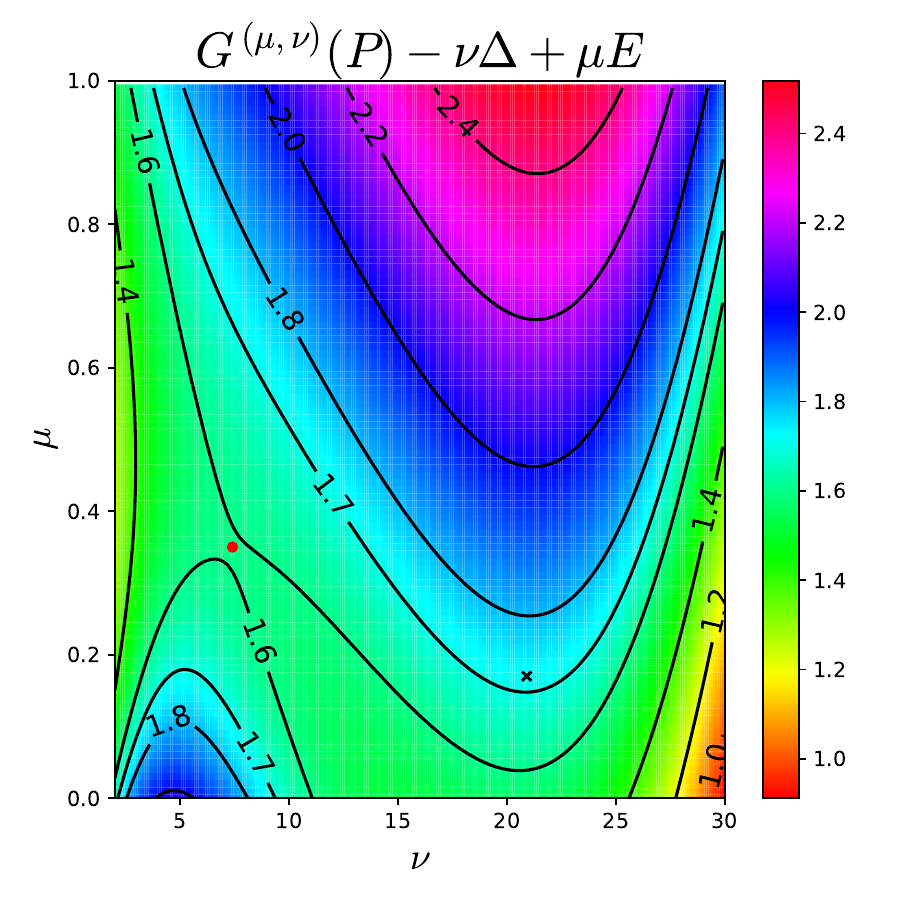}
        \subcaption{$E=1.0$}\label{fig:non_convexity_mu_nu_right}
    \end{minipage}
    \caption{$G^{(\mu, \nu)}(P)
- \nu \Delta + \mu E $ for \#1 of Table~\ref{table1}. The distortion measure is given by (\ref{distortion}) }
    \label{fig:non_convexity_mu_nu}
\end{figure}

Here, we demonstrate the non-convexity of 
\begin{align}
G^{(\mu, \nu)}(P) 
= 
\min_{p_Y \in \mathcal{P(Y)}} G^{(\mu, \nu)}(p_Y|P). 
\end{align}
with respect to $\mu$ and $\nu$.
From (\ref{eq.theorem2}) and (\ref{def:R_B}), we have
$
R_{\rm M}(E| \Delta, P ) =
\sup_{\nu\geq 0} 
\inf_{\mu \geq 0}
\{
G^{(\mu, \nu)}(P)
- \nu \Delta + \mu E 
\}$ and
$
R_{\rm B}(E| \Delta, P ) =
\inf_{\mu \geq 0}
\sup_{\nu\geq 0} 
\{
G^{(\mu, \nu)}(P)
- \nu \Delta + \mu E 
\}$.
Two examples of the heat map and the contour plot of $G^{(\mu, \nu)}(P) - \nu \Delta + \mu E $ are shown in Fig.~\ref{fig:non_convexity_mu_nu}, where the parameters are given by \#1 of Table~\ref{table1} and $E$ is $0.5$ (left) and $1.0$ (right). 
Note that the difference between these two figures is the last term $\mu E$. The optimal $(\mu, \nu)$ for $R_{\rm M}(E|\Delta, P)$ is shown by a point marker and that for $R_{\rm B}(E|\Delta, P)$ is shown by a "x" marker. 
These graphs visually show the non-convexity of the function $G^{(\mu, \nu)}(P)$. 

For $E=0.5$, the optimal $(\mu, \nu)$ for $R_{\rm M}(E|\Delta, P)$ and $R_{\rm B}(E|\Delta, P)$ are $(0, 20.4)$ and $(0.17, 20.9)$, respectively. Although the figure for $E=0.2$ is not shown here, the optimal parameters are the same for $E=0.2$.
The inverse of the Marton exponent is flat because the optimal $\mu$ is zero, while the inverse of the Blahut exponent is a straight line between $E=0.2$ and $0.5$ with a slope of $\mu=0.17$.
For $E=1.0$, on the other hand, the optimal point of $(\mu, \nu)$ for the inverse of the Marton exponent is $(0.35, 7.4)$, which is taken at an interior point of the feasible region. This point is a saddle point of $G^{(\mu, \nu)}(P) - \nu \Delta + \mu E $. 
It is important to note that for computing the Blahut exponent (\ref{def:E_B}), we first maximize 
$G^{(\mu, \nu)}(P)-\nu \Delta$ with respect to $\nu$ for fixed $\mu=1/\rho$.
Fig.~\ref{fig:non_convexity_mu_nu} (b) shows that when $\mu$ is fixed, for example $\mu=0.2$,
this function has two local maxima. Therefore, even when computing the Blahut exponent, we must be careful about the possible existence of local maxima that are not global.

\ifISIT
\else 

\begin{figure}
    \centering
    \includegraphics[width=0.4\textwidth]{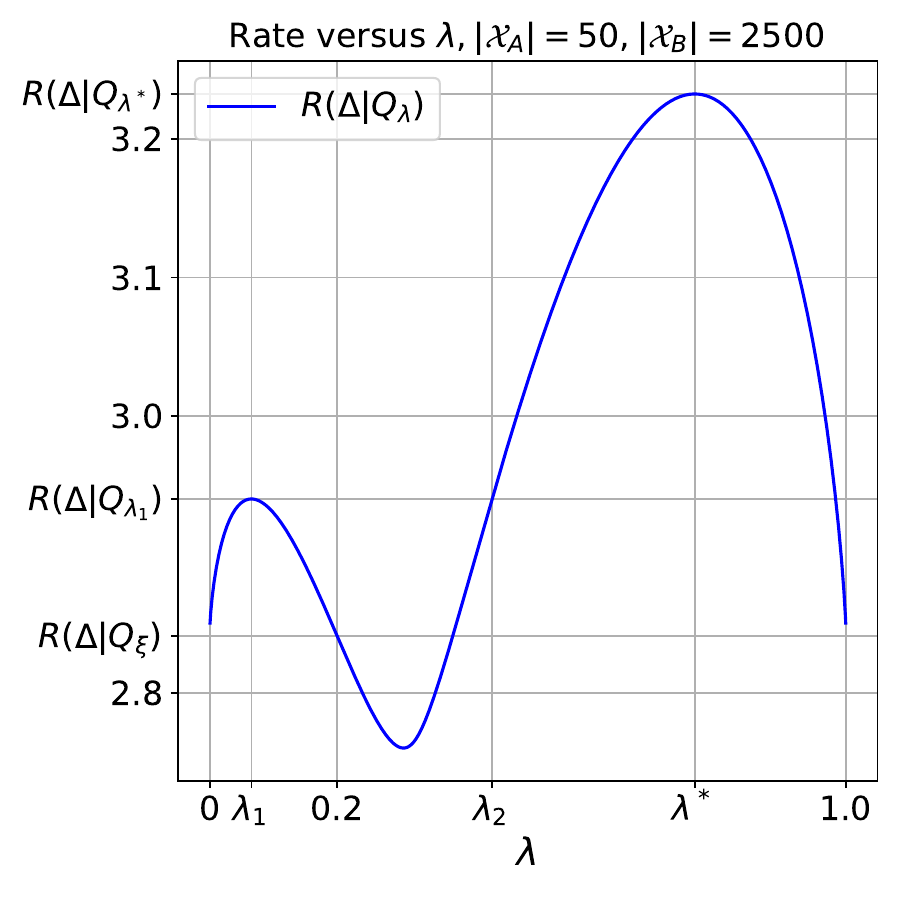}
    \caption{The rate-distortion function for \#2 of Table~\ref{table1}}
    \label{fig:rate_distortion2}
\end{figure}

\begin{figure}
    \centering
    \includegraphics[width=0.4\textwidth]{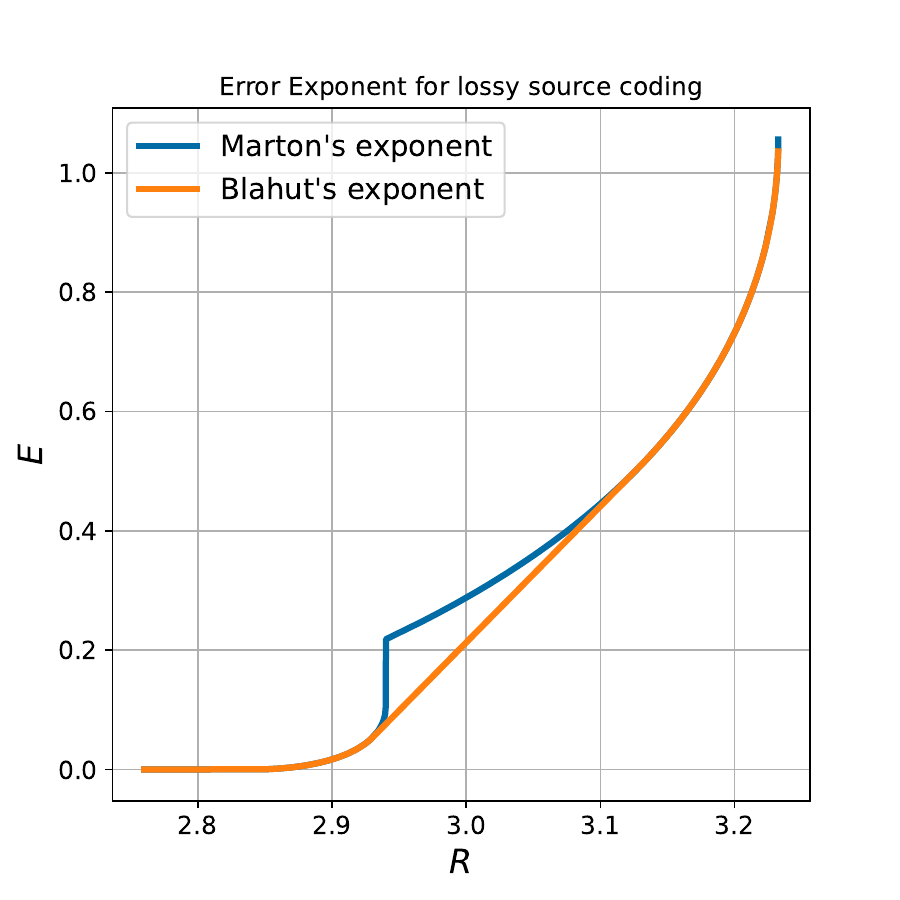}
    \caption{ Error exponents for \#2 of Table~\ref{table1}}
    \label{fig:error_exponent2}
\end{figure}

Next, we use \#2 listed in Table~\ref{table1} to see the non-convexity of the rate-distortion function and the Marton exponent function.
The rate-distortion function for this case is shown in Fig.~\ref{fig:rate_distortion2}. 
The global maximum is found at $\lambda^* = 0.762$ and a local maximum at $\lambda=\lambda_1=0.065$.
Then, the rate-distortion function of this case was computed by the Arimoto-Blahut algorithm. 
Marton's and Blahut's exponents are shown in Fig.~\ref{fig:error_exponent2}, where $P=Q_\xi$ with $\xi = 0.2$. 
We observe that Marton's exponent jumps from $D(Q_{\lambda_1}\|Q_{0.2})=0.103$ to $D(Q_{\lambda_2}\|Q_{0.2})=0.220$ at $R=R(\Delta|Q_{\lambda_1}) = R(\Delta|Q_{\lambda_2}) = 2.940$. 
In Fig.~\ref{fig:inverse2}, $R_{\rm M}(E|\Delta,P)$ computed by the proposed method is drawn. 
We confirm that the graph is correctly computed.

\begin{figure}
    \centering
    \includegraphics[width=0.4\textwidth]{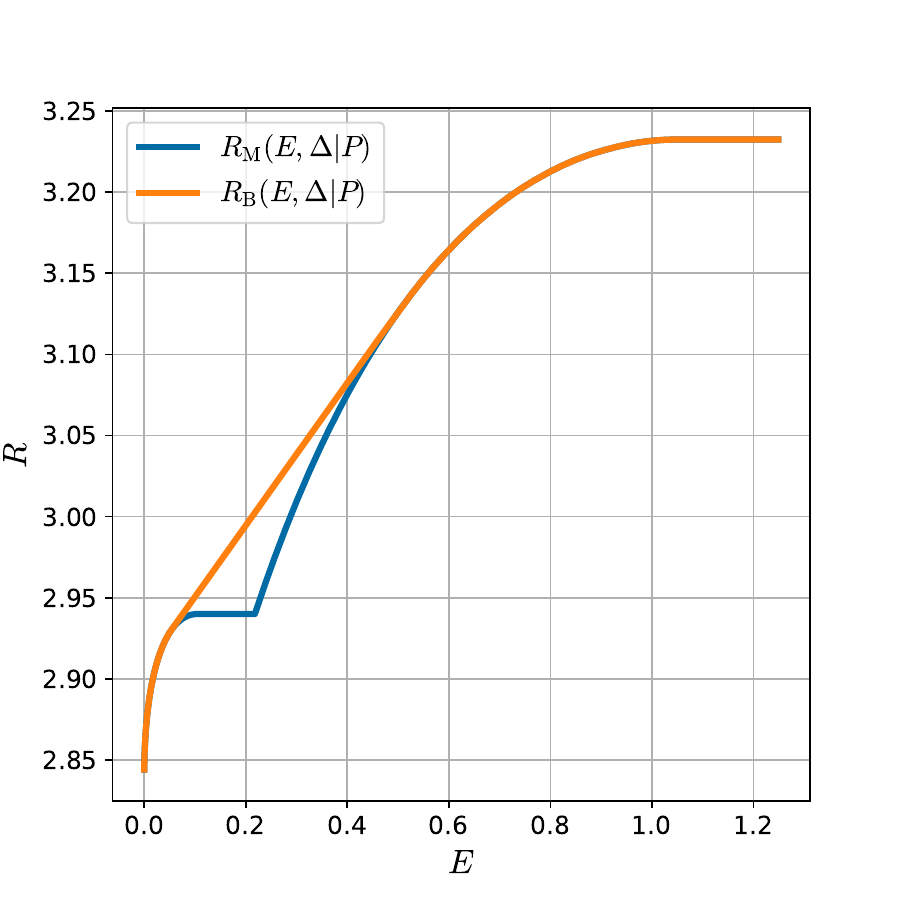}
    \caption{$R_{\rm M}(E | \Delta, P)$ for \#2 of Table~\ref{table1} }
    \label{fig:inverse2}
\end{figure}

\fi

\ifISIT

\else



\section{Proof of Theorem~\ref{theorem2}}
\label{section:proof:theorem2}
In this section, we give the proof of Theorem~\ref{theorem2}, which consists of three steps.
The first step is to use a parametric expression for the rate-distortion function. Since this expression formula is well known, it is natural to examine what can be done with it.
The second step is to use the minimax theorem. The effect of this is to swap the order of the outer conditional maximization problem and the inner minimization problem for the probability distribution over $\mathcal{Y}$. Without this step, the Lagrange undetermined multiplier method cannot be successfully applied to the conditions of the outer conditional maximization problem.
The third step introduces the Lagrange multiplier for the conditional maximization problem. This yields a representation with two Lagrange multipliers $(\mu,\nu)$. Only the case $\mu=0$ requires special processing, but otherwise the method is standard.

\subsubsection{A parametric expression for the rate-distortion function}
The function $R_{\rm M}(E|\Delta, P)$ is much easier to analyze than $E_{\rm M}(R|\Delta, P)$ because 
the feasible region in the right-hand side of (\ref{R_M}) is convex.
In (\ref{R_M}), however, the objective function is the rate-distortion function, which is not necessarily convex.
To circumvent this issue, we use the following parametric expression of $R(\Delta|q_X)$. 
This is the first step. 
\begin{lemma}
\label{lemma4}
We have
\begin{align}
& R(\Delta | q_X) =
\sup_{\nu \geq 0}
\Big[ -\nu \Delta 
+ \min_{ p_Y \in \mathcal{P(Y)} }
-\sum_x q_X(x) \log \sum_y p_Y(y) \mathrm{e}^{-\nu d(x,y)}
\Big]. \label{parametric_Rate_distortion}
\end{align}
\end{lemma}

One can refer~\cite[Corollary 8.5]{Csiszar-KornerBook} for the proof. 
\ifISIT
\else
\fi

We should mention that (\ref{parametric_Rate_distortion}) is related to an important notion called the $d$-tilted information density~\cite{Kostina2012}, although this relation is not used in this paper. 
Denote the $\nu$ and $p_Y$ that attains (\ref{parametric_Rate_distortion}) by $\nu^*$ and $p_Y^*$.
Then, 
\begin{align}
\jmath_X(x,d) := -\log \sum_y p_Y^*(y) {\rm e}^{-\nu^* (d(x,y)-\Delta)}
\end{align}
is called $d$-tilted information density and we observe that 
$R(\Delta|q_X) = {\rm E}_{q_X} [ \jmath_X(X,d) ]$ holds.

\subsubsection{Minimax theorem}
We substitute (\ref{parametric_Rate_distortion}) into (\ref{R_M}). 
Then, except for the maximization over $\nu\ge 0$, 
we have to evaluate the quantity defined by
    \begin{align}
    \max_{
\genfrac{}{}{0pt}{}{
q_X \in \mathcal{P(X)}:
}{
D(q_X\|P) \leq E
}
} 
\min_{ p_Y \in \mathcal{P(Y)} }
-\sum_x q_X(x) \log \sum_y p_Y(y) \mathrm{e}^{-\nu d(x,y)}. 
    \label{saddlepoint}
\end{align}

The second step is the exchange of the order of max and min
in (\ref{saddlepoint}). 
For deriving an algorithm for computing $R_{\rm M}(E|\Delta,P)$, 
the saddle point (\ref{saddlepoint}) should be transformed into minimization or maximization problems. 
In order to derive such an expression, we exchange the order of maximization w.r.t.~$q_X$ and minimization w.r.t.~$p_Y$.
The following lemma is essential for deriving the exact parametric expression for the inverse function of the error exponent.
\begin{lemma} \label{application_of_minimax_theorem}
    For any $E\ge 0$ and $\nu\ge 0$, we have
    \begin{align}
    & 
    \max_{
    \genfrac{}{}{0pt}{}{ q_X \in \mathcal{P(X)}: } 
    { D(q_X\|P) \leq E }
    } 
\min_{ p_Y \in \mathcal{P(Y)} }
-\sum_x q_X(x) \log \sum_y p_Y(y) \mathrm{e}^{-\nu d(x,y)}
 \notag\\
    &=
    \min_{p_Y \in \mathcal{P(Y)}} 
    \max_{
    \genfrac{}{}{0pt}{}{q_X \in \mathcal{P(X)}: }{ D(q_X\|P) \leq E } 
    }
    -\sum_x q_X(x) \log \sum_y p_Y(y) {\rm e}^{-\nu d(x,y) } 
    \end{align}
\end{lemma}

The validity of this exchange relies on Sion's minimax theorem~\cite{Sion1958}.

\begin{theorem}[Sion~\cite{Sion1958}] \label{MinimaxTheorem}
Let $\mathcal{P}$ and $\mathcal{Q}$ be convex, compact spaces, and $f(p, q)$ be a function on $\mathcal{P\times Q}$. 
If $f(p, q)$ is lower semicontinuous and quasi-convex on $p \in \mathcal{P}$
for any fixed $q\in\mathcal{Q}$
and $f(p, q)$ is upper semicontinuous and quasi-concave in $q \in \mathcal{Q}$
for any fixed $p\in\mathcal{P}$, then
\begin{align}
    \inf_{ p \in \mathcal{P} } \sup_{ q \in \mathcal{Q} } f(p,q) = 
    \sup_{ q \in \mathcal{Q} } \inf_{ p \in \mathcal{P} } f(p,q) .
\end{align}
\end{theorem}

\textit{Proof of Lemma~\ref{application_of_minimax_theorem}}: 
As stated above, the objective function of (\ref{saddlepoint}) is linear in $q_X$ and convex in $p_Y$. 
The set $\{ q_X\in \mathcal{P(X)}: D(q_X\|P) \leq E\}$ is convex because $D(q_X\|P)$ is convex in $q_X$. 
Hence, we can apply Theorem~\ref{MinimaxTheorem} to (\ref{saddlepoint}). 
\hfill\IEEEQED

\subsubsection{The second Lagrange multiplier}
Next, we define the following functions:
\begin{definition}
\label{def:Gmunu}
For $\mu\geq0, \nu\geq0$, $p_Y\in \mathcal{P(Y)}$,
and $P\in \mathcal{P(X)}$, we define 
\begin{align}
G^{(\nu)}(E, p_Y |P)
&= 
    \max_{
    \genfrac{}{}{0pt}{}{q_X \in \mathcal{P(X)}: }{ D(q_X\|P) \leq E } 
    }
    -\sum_x q_X(x) \log \sum_y p_Y(y) {\rm e}^{-\nu d(x,y) }, 
    \label{GnuEp_YP}
    \\
G^{(\mu,\nu)}(p_Y|P) 
&= \max_{q_X\in \mathcal{P(Y)}} \bigg[
- \mu D( q_X \| P )
-\sum_{x} q_X(x) \log \sum_y p_Y(y) \mathrm{e}^{-\nu d(x,y)} \bigg], 
\label{Gmunup_YP}
\\
G^{(\mu, \nu)}(P) & = \min_{p_Y \in \mathcal{P(Y)}} G^{(\mu, \nu)}(p_Y|P). 
\label{GmunuP}
\end{align}
\end{definition}

The last step is to transform (\ref{GnuEp_YP}), which is a constrained maximization, into an unconstrained maximization by introducing a Lagrange multiplier. For this purpose, we have defined (\ref{Gmunup_YP}). Then,  (\ref{Gmunup_YP}) is explicitly obtained as follows:
\begin{lemma}
    \label{Lemma_G_nu}
For $\mu,\nu\geq 0$, $p_Y\in \mathcal{P(Y)}$,
and $P\in \mathcal{P(X)}$, we have
\begin{align}
G^{(\mu,\nu)} ( p_Y | P) &=
\begin{cases}
\mu \log \sum_x P(x) \left\{ \sum_{y} p_Y (y) \mathrm{e}^{-\nu d(x,y)} \right\}^{-1/\mu} & \text{ if } \mu > 0,\\
- \log \min_x \sum_y p_Y(y) \mathrm{e}^{-\nu d(x,y)} &\text{ if } \mu = 0.
\end{cases}  
\label{Gmunup_YP2}
\end{align}
\end{lemma}

We have the following lemma.
\begin{lemma}
    \label{lemma3}
    For $\nu\geq 0,E\geq 0, p_Y\in \mathcal{P(Y)}$, and $P\in\mathcal{P(X)}$, we have
\begin{align}
G^{(\nu)}(E, p_Y|P) = \inf_{\mu\ge 0} \{ \mu E + G^{(\mu,\nu)} (p_Y | P) \}. \label{G_nu2}
\end{align}
\end{lemma}
\ifISIT
See ~\cite{YutakaISIT2023arXiv} for the proofs of Lemmas~\ref{Lemma_G_nu} and \ref{lemma3}.
\else
The proofs of Lemmas~\ref{Lemma_G_nu} and~\ref{lemma3} appear in Appendix C and D. 
\fi

Eq.~(\ref{G_nu2}) is a parametric expression for (\ref{GnuEp_YP}).

\textit{Proof of Theorem~\ref{theorem2}:}
We have 
\begin{align}
    & R_{\rm M}(E|\Delta, P) \notag\\
    & 
    \stackrel{\rm (a)}
    = 
    \max_{
    \genfrac{}{}{0pt}{}{ q_X \in \mathcal{P(X)}: } 
    { D(q_X\|P) \leq E }
    }
    \sup_{\nu \geq 0}
\Big[ -\nu \Delta 
+ 
\min_{ p_Y \in \mathcal{P(Y)} }
-\sum_x q_X(x) \log \sum_y p_Y(y) \mathrm{e}^{-\nu d(x,y)}
\Big] \notag\\
    & 
    \stackrel{\rm (b)}
    = 
    \sup_{\nu \geq 0}
    \min_{ p_Y \in \mathcal{P(Y)} }
    \max_{
    \genfrac{}{}{0pt}{}{ q_X \in \mathcal{P(X)}: } 
    { D(q_X\|P) \leq E }
    }
    \Big[ -\nu \Delta -\sum_x q_X(x) \log \sum_y p_Y(y) \mathrm{e}^{-\nu d(x,y)}
\Big] \notag\\
    & 
    \stackrel{\rm (c)}
    = 
    \sup_{\nu \geq 0}
    \min_{ p_Y \in \mathcal{P(Y)} }
    \Big[ -\nu \Delta + G^{(\nu)}(E, p_Y|P) 
\Big] \notag\\
    &
    \stackrel{\rm (d)}
    = 
    \sup_{ \nu \ge 0 }
    \min_{ p_Y \in \mathcal{P(Y)} }
    \Big[ -\nu \Delta + \inf_{ \mu \ge 0 }  \big\{ \mu E + 
    G^{(\mu, \nu)} (q_Y|P) \big\}
    \Big] \notag\\ 
    &
    \stackrel{\rm (e)}
    = 
    \sup_{ \nu \ge 0 }
    \inf_{ \mu \ge 0 }
    \min_{ p_Y \in \mathcal{P(Y)} }
    \Big[ -\nu \Delta + \mu E + G^{(\mu, \nu)} (q_Y|P)
\Big] \notag\\ 
    &
    \stackrel{\rm (f)}
    = 
    \sup_{ \nu \ge 0 }
    \inf_{ \mu \ge 0 }
    \Big[ -\nu \Delta + \mu E + G^{(\mu, \nu)} (P)
\Big]. \label{parametric_R_M_E}
\end{align}
Step (a) follows from Lemma~\ref{lemma4},
Step (b) follows from Lemma~\ref{application_of_minimax_theorem},
Step (c) follows from Eq.~(\ref{GnuEp_YP}),  
Step (d) follows from Eq.~(\ref{G_nu2}), 
Step (e) holds because the inf and min operations are interchangeable, 
and Step (f) follows from Eq.~(\ref{GmunuP}). \hfill\IEEEQED

\ifISIT
\else 

\begin{remark}
The minimax theorem in Lemma~\ref{application_of_minimax_theorem} was the key step and 
the order of the above four steps is of crucial importance. 
The straightforward Lagrange dual of the expression (\ref{R_M}) is defined as   
\begin{align}
\widetilde R_{\rm M}(E|\Delta, P) 
&=
\inf_{\mu\ge 0} \left\{ 
\max_{ q_X \in \mathcal{P(X)}:}
[R(\Delta | q_X) - \mu D(q_X\|P) ]
+ \mu E
\right\}. \label{R_M_hat}
\end{align}
However, applying the same steps in the proof of Theorem~\ref{theorem2} to the expression (\ref{R_M_hat}), we see that $\widetilde R_{\rm M}(E|\Delta, P) $ coincides with $R_{\rm B}(E|\Delta, P) $. Therefore the expression (\ref{R_M_hat}) is also suboptimal. It means that it was very important to apply Lemma~\ref{lemma4} first to avoid non-convexity of the rate-distortion function.

\end{remark}

\fi

\section*{Acknowledgments}
The author thanks Professor Yasutada Oohama, Dr. Yuta Sakai, and Shun Watanabe for their valuable comments. 

\bibliographystyle{IEEEtran}



\appendix

\section{Proofs}
We give the proofs of Theorem~\ref{theorem_inverse_Blahut} and Theorem~\ref{theorem1} in Appendices A and B.
The proof of Property~\ref{property_R_M}d) is given in Appendix C and the proofs of Lemmas~\ref{lemma.1}, \ref{Lemma_G_nu} and \ref{lemma3} are given in Appendices D, E, and F. 
The proofs of the lemmas required for the proof of Property~\ref{property_R_M}d) are given in Appendices G, H, and I.

\subsection{Proof of Theorem~\ref{theorem_inverse_Blahut}}
\label{appendix_proof_theorem_inverseBlahut}
\begin{IEEEproof}
From (\ref{def:E_B}), we define
\begin{align}
R^{(\rho)}(\Delta, P)
=
\frac{-1}{\rho}
\inf_{\nu\geq 0} \left[
\rho \nu \Delta 
-
\min_{p_Y \in \mathcal{P(Y)}} 
\log \sum_{x \in \mathcal{X} }
P(x) 
\left(
\sum_{y\in \mathcal{Y}} p_Y(y) {\rm e}^{-\nu d(x,y) }
\right)^{-\rho}
\right]. \label{eq.39}
\end{align}
so that $ E_{\rm B}(R|\Delta, P)
= \sup_{\rho\ge 0} [ \rho R - \rho R^{(\rho)}(\Delta, P)]$ holds. 
Then, $ \rho R - \rho R^{(\rho)}(\Delta, P)$ is considered as a tangent line with slope $\rho$ and $R^{(\rho)}(\Delta, P)$ is the $R$-axis intercept of this tangent line. For the inverse function, the slope parameter is $\mu = 1/\rho$. 
This interpretation implies that the inverse of $E_{\rm B}(R | \Delta, P)$ is defined as 
\begin{align}
    R_{\rm B}(E|\Delta, P)
    = 
    \inf_{\mu \ge 0} \{ 
    \mu E + R^{(1/\mu)}(\Delta, P) 
    \}. \label{eq.40}
\end{align}
Substituting (\ref{eq.40}) into (\ref{eq.39}), we have
\begin{align}
    R_{\rm B}(E|\Delta, P)
    &= 
    \inf_{\mu \ge 0} 
    \left\{ \mu E - 
    \inf_{\nu\ge 0} \left[ 
    \nu
    \Delta
    +\min_{p_Y \in \mathcal{P(Y)}} 
    \mu 
    \log \sum_{x \in \mathcal{X} } P(x) 
\left(
\sum_{y\in \mathcal{Y}} p_Y(y) {\rm e}^{-\nu d(x,y) }
\right)^{-1/\mu}
    \right] \right\}\notag\\
    &= 
        \inf_{\mu \ge 0} 
    \sup_{\nu\ge 0}
    \left[ \mu E - \nu \Delta
    +\min_{p_Y}
    G^{(\mu,\nu)}(p_Y|P)
    \right]
\end{align}
This completes the proof.
\end{IEEEproof}
\subsection{Proof of Theorem~\ref{theorem1}}
We use the following lemma Ahlswede~\cite{Ahlswede1990} to prove Theorem~\ref{theorem1}.
\begin{lemma}
\label{lemma1}
For any $P\in \mathcal{P(X)}$ with $\mathcal{X}=\mathcal{X}_A \cup \mathcal{X}_B$ where $\mathcal{X}_A$ and $\mathcal{X_B}$ are disjoint, 
define $\lambda=  \sum_{x\in \mathcal{X}_A} P(x) $. We have
\begin{align}
R(\Delta | \lambda Q_A + (1 - \lambda) Q_B ) \geq R(\Delta |P).
\end{align}
\end{lemma}
See~\cite{Ahlswede1990} for the proof.

\textit{Proof of Theorem~\ref{theorem1}:}
Let $q_X^* \in \mathcal{P(X)}$ be an optimal distribution that attains 
$E_{\rm M}(R | \Delta , Q_{\xi}) = \min_{q_X: R(\Delta|q_X) \geq R}  D(q_X\|Q_{\xi}).$ 
Put $\lambda^* = \sum_{x\in \mathcal{X}_A } q_X^*(x)$.
We will show that $q_X^*$ is expressed by $\lambda^* Q_A + (1-\lambda^*) Q_B$. 

Let $q_A^*(x) = q_X^*(x)/\lambda^*$ for $x\in \mathcal{X}_A$ and
$q_B^*(x)= q_X^*(x)/(1-\lambda^*)$ for $x\in \mathcal{X}_B$.
Then, we have
\begin{align}
& D(q_X^* \| Q_{\xi} ) \notag\\
& =
  \sum_{x\in \mathcal{X} } q_X^{*}(x) \log \frac{q_X^*(x)}{ Q_{\xi} (x) }  \notag \\
&=
    \sum_{x\in \mathcal{X}_A } \lambda^* q_A^{*}(x) \log \frac{ \lambda^* q_A^{*}(x) }{ \frac{ \xi }{ |\mathcal{X}_A| } } + 
    \sum_{x\in \mathcal{X}_B } (1-\lambda^*) q_B^{*}(x) \log \frac{ (1-\lambda^*) q_B^*(x)}{ \frac{1-\xi}{ |\mathcal{X}_B| }  } 
\notag\\
&=
\lambda^* \left \{ \log \frac{ \lambda^* | \mathcal{X}_A | }{\xi}
+ \sum_{x\in\mathcal{X}_A} q_A^*(x) \log q_A^*(x) \right \} \notag\\
& \quad+
(1-\lambda^*) \left\{  \log \frac{(1-\lambda^*) | \mathcal{X}_B | }{1-\xi}
+ \sum_{x\in\mathcal{X}_B} q_B^*(x) \log q_B^*(x) \right\} \notag\\
&
\stackrel{(a)}
\geq
\lambda^* \left \{ \log \frac{ \lambda^* | \mathcal{X}_A | }{\xi}
-\log |\mathcal{X}_A| \right \} +
(1-\lambda^*) \left\{  \log \frac{(1-\lambda^*) | \mathcal{X}_B | }{1-\xi}
-\log |\mathcal{X}_B| \right\} \notag\\
&=D_2(\lambda^* \| \xi) = D ( \lambda^* Q_A + ( 1-\lambda^*) Q_B \| Q_{\xi} ),
\label{inequality}
\end{align}
Equality in (a) holds if and only if $q_A^*(x) = 1/|\mathcal{X}_A|$
 and $q_B^*(x) = 1/|\mathcal{X}_B|$. 
From Lemma~\ref{lemma1}, we have 
$ R(\Delta | \lambda^* Q_A + (1-\lambda^*) Q_B) \geq R(\Delta | q_X^*) $ ($\geq R$).
Therefore $ \lambda^* Q_A + ( 1-\lambda^*) Q_B$ is included in the feasible region
$\{q_X: R(\Delta|q_X) \ge R\}$. 
Since we assumed $q_X^*$ is optimal, we must have $q_X^* = \lambda^* Q_A + ( 1-\lambda^*) Q_B$. 
This completes the proof.\hfill$\IEEEQED$

\subsection{Proof of Property~\ref{property_R_M}d).}
\label{sec:appendix:continuity}
The continuity property of $R_{\rm M}(E|\Delta, P)$ is stated in Lemma 2 of~\cite{Haroutunian2000}. 
However, the proof is not clear.
In this section, we give a detailed proof of the continuity of $R_{\rm M}(E | \Delta, P)$. 

In order to prove the Property~\ref{property_R_M}d),
we divide optimal distributions $q_X^*$ of the optimization problem $R_{\rm M}(E|\Delta, P)$ defined by (\ref{R_M}) into two cases: it is achieved at the boundary or it is achieved at an interior point of the feasible region.
Here, we divide the optimal distribution $q_X^*$ of the optimization problem $R_{\rm M}(E|\Delta, P)$ defined by (\ref{R_M}) into two cases: one is achieved at the boundary and the other at an interior point of the feasible region.
The optimal distribution $q_X^*$ satisfies 
$D(q_X^* \| P)=E$ in the former case and $D(q_X^* \| P)<E$ in the latter case. 
To distinguish between these two cases more clearly, we introduce the following function.
\begin{align}
    \widetilde R_{\rm M}(E|\Delta,P) 
     = 
     \max_{ 
     \genfrac{}{}{0pt}{}{q_X \in \mathcal{P(X)}:}{D(q_X\|P) = E}
     } 
    R(\Delta | q_X).
\end{align}
For avoiding the set $\{q_X \in \mathcal{P(X)}: D(q_X\|P) = E\}$ to be empty,
$\widetilde R_{\rm M}(E|\Delta,P)$ is defined for $E\in [0, E^*]$, where 
$E^* = -\log \min_{ \genfrac{}{}{0pt }{}{x\in \mathcal{X}:}{P(x)>0} } P(x) $.

We have the following lemma.
\begin{lemma}\label{lemma:Rtilde}
    For $E\ge 0$, we have
    \begin{align}
        R_{\rm M} (E|\Delta, P) = \max_{E'\le E} \widetilde R_{\rm M}(E' |\Delta,P) .
        \label{eqn.lemma.R_tilde}
    \end{align}
\end{lemma}
See Appendix~\ref{Section:Proof:lemma:Rtilde} for the proof.

We prove the continuity of $\widetilde R_{\rm M}(E | \Delta, P)$, which directly implies the continuity of $R_{\rm M}(E | \Delta, P)$. 
Before stating the proof, we give the following two lemmas necessary for the proof.
They are necessary to construct a delta-epsilon proof for Property~\ref{property_R_M}d).

\begin{lemma}
\label{lemmaB}
    For any $\epsilon>0$ and any $p_X \in \mathcal{P(X)}$,
    there exists a $\xi=\xi(\epsilon)$ such that for any $q_X \in \mathcal{P(X)}$ with $\lVert p_X - q_X \rVert \le \xi(\epsilon)$ satisfies
    $\lvert R(\Delta|p_X) - R(\Delta|q_X) \rvert \le \epsilon$.
    This implies that for any $\epsilon>0$ and $p_X\in \mathcal{P(X)}$, 
    there exists a $\xi=\xi(\epsilon)>0$ such that 
    the following two inequalities hold.
    \begin{align}
        \max_{ \genfrac{}{}{0pt}{}{q_X \in \mathcal{P(X)}:}{ \lVert p_X - q_X \rVert_2 \le \xi } } 
        R(\Delta|q_X ) \le R(\Delta| p_X ) + \epsilon, \label{lemmaBmax} \\
        \min_{ \genfrac{}{}{0pt}{}{ q_X \in \mathcal{P(X)}:}{ \lVert p_X -q_X  \rVert_2 \le \xi } } 
        R(\Delta| q_X ) \ge R(\Delta| p_X ) - \epsilon. \label{lemmaBmin} 
    \end{align}
\end{lemma}

\begin{lemma}
    \label{lemmaA}
    For any $E\in [0, E^*]$ and any $\xi>0$, there exists a $\delta(E,\xi)>0$ such that
    for any $p_X \in \mathcal{P(X)}$ satisfying
    \begin{align} 
    E - \delta(E,\xi)
      \le D(p_X\|P) \le E + \delta(E,\xi),
    \end{align}
    there exists a $q_X\in \mathcal{P(X)}$ 
    satisfying
    \begin{align}
      D(q_X\|P) &= E,\\
    \lVert p_X - q_X \rVert_2 &\le \xi.
    \end{align}
\end{lemma}

The proofs for Lemmas~\ref{lemmaB} and~\ref{lemmaA} are given in Appendix F and G.

\textit{Proof of Property~\ref{property_R_M}d):}
We first prove the continuity of $\widetilde R ( E | \Delta, P)$ in $E$. 
By Lemma~\ref{lemmaB}, for any $\epsilon>0$, there exists a $\xi=\xi(\epsilon)$ such that
any $p_X \in \mathcal{P(X)}$ satisfies (\ref{lemmaBmax}) and (\ref{lemmaBmin}).
Fix this $\xi(\epsilon)$.
For this choice of $\xi(\epsilon)$ and any $E\in [0,E_*]$, from Lemma~\ref{lemmaA}
there exists a $\delta = \delta(E,\xi(\epsilon))>0$ such that for any $p_X\in \mathcal{P(X)}$ satisfying
$|D(p_X\|P)-E|\le \delta$ there exists a $q_X\in \mathcal{P(X)}$ satisfying $D(q_X\|P)=E$ and  
$\lVert p_X-q_X\rVert_2 \le \xi(\epsilon)$.
Then, for any for any $E' \in [E - \delta, E + \delta]$, 
the following chain of inequalities holds. 
\begin{align}
    & \widetilde R(E' | \Delta, P) \notag \\
    & = \max_{
    \genfrac{}{}{0pt}{}{p_X\in \mathcal{P(X)} }{ D(p_X\|P) = E' }} 
    R(\Delta | p_X) \notag\\
    & \stackrel{\rm (a) }\le
    \max_{
    \genfrac{}{}{0pt}{}{ p_X\in \mathcal{P(X)} }{ D(p_X\|P) = E' }} 
    \min_{
    \genfrac{}{}{0pt}{}{ q_X\in \mathcal{P(X)} }{ \lVert p_X - q_X \rVert_2 \le \xi(\epsilon) }
    }
    R(\Delta | q_X) +\epsilon \notag\\
    & \stackrel{\rm (b) }\le
    \max_{
    \substack{ p_X \in \mathcal{P(X)} \\ D(p_X\|P) = E'}
    }
    \min_{
    \substack{ q_X \in \mathcal{P(X)}: \\ 
            \lVert p_X - q_X \rVert_2 \le \xi(\epsilon) \\ 
            D(q_X\|P) = E}
    }
    R(\Delta | q_X) +\epsilon \notag\\
    & \stackrel{\rm (c) }\le
    \max_{
    \substack{ p_X \in \mathcal{P(X)} \\ D(p_X\|P) = E'}
    }
    \min_{
    \substack{ q_X \in \mathcal{P(X)}: \\ 
            \lVert p_X - q_X \rVert_2 \le \xi(\epsilon) \\ 
            D(q_X\|P) = E}
    }
    \widetilde{R}(E|\Delta , P) +\epsilon \notag\\
    & 
    \stackrel{\rm (d) }    
    = \widetilde{R}(E|\Delta, P) +\epsilon \label{proof.continuity.eq1}
\end{align}
Step (a) follows from (\ref{lemmaBmin}).
Step (b) holds because additional constraint on the minimization increases the minimum value.
Step (c) follows from that the maximum value of 
$R(\Delta|q_X)$ under the condition $D(q_X\|P)=E$ and a fixed $\Delta$ is given by $\tilde R(E|\Delta, P)$. 
Step (d) holds because 
for any $p_X$ satisfying $D(p_X||P)=E'$,
there exists $q_X\in \mathcal{P(X)}$ satisfying
$D(p_X\|P) = E'$, $D(q_X\|P) = E$,  and 
$\lVert p_X - q_X \rVert_2 \le \xi(\epsilon)$ due to Lemma~\ref{lemmaA}.

On the other hand, for any $ E' \in [E - \delta, E + \delta] $, we have the following chain of inequalities.
\begin{align}
    & \widetilde R(E' | \Delta, P) \notag \\
    & =\max_{
    \genfrac{}{}{0pt}{}{p_X\in \mathcal{P(X)} }{ D(p_X\|P) = E' }} 
    R(\Delta | p_X) \notag\\
    & \stackrel{\rm (a) }\ge
    \max_{
    \genfrac{}{}{0pt}{}{ p_X\in \mathcal{P(X)} }{ D(p_X\|P) = E' }} 
    \max_{
    \genfrac{}{}{0pt}{}{ q_X\in \mathcal{P(X)} }{ \lVert p_X - q_X \rVert \le \xi(\epsilon) }
    }
    R(\Delta | q_X) -\epsilon \notag\\
    & \stackrel{\rm (b) }\ge
    \max_{
    \substack{ p_X \in \mathcal{P(X)} \\ D(p_X\|P) = E'}
    }
    \max_{
    \substack{ q_X \in \mathcal{P(X)}: \\ 
            \lVert p_X - q_X \rVert \le \xi(\epsilon) \\ 
            D(q_X\|P) = E}
    }
    R(\Delta | q_X) - \epsilon \notag\\
    & \stackrel{\rm (c) } =
    \max_{
    \substack{ q_X \in \mathcal{P(X)}: \\ 
            D(q_X\|P) = E}
    }
    R(\Delta | q_X) -\epsilon \notag\\
    & 
    = \widetilde{R}(E| \Delta , P) -\epsilon  \label{proof.continuity.eq2}
\end{align}
Step (a) follows from (\ref{lemmaBmax}).
Step (b) holds because additional constraint on the maximization decreases the maximum value.
Step (c) follows from that, by Lemma~\ref{lemmaA}, for any $p_X$ satisfying $D(p_X\|P)=E'$,
there exists $q_X$ such that $D(q_X\|P)=E$ and $\lVert p_X-q_X\rVert\le \xi(\epsilon)$ hold
and that the objective function $R(\Delta|q_X)$ is independent of $p_X$.
Combining (\ref{proof.continuity.eq1}) and (\ref{proof.continuity.eq2}), we have 
\begin{align}
    \lvert \widetilde R(E|\Delta, P ) - \widetilde R(E'|\Delta, P) \rvert < \epsilon.
\end{align}
This proves the continuity of $\widetilde R(E|\Delta, P)$.
The continuity of $R(E|\Delta, P )$ follows directly from (\ref{eqn.lemma.R_tilde}), 
which completes the proof.\hfill\IEEEQED


\subsection{Proof of Lemma~\ref{lemma.1}}
Let 
$\rho$ be any non-negative number
and
$q_X^*$ be an optimal distribution that achieves
$ E_{\rm M}(R|\Delta, P)  = \min_{q_X: R(\Delta|q_X)\geq R} D(q_X\|P) $. 
Then, we have 
\begin{align}
    & E_{\rm M}(R|\Delta, P) = D(q_X^*\|P) \notag\\
    & \stackrel{\rm (a)}
      \ge \{ D(q_X^*\|P) -\rho[R(\Delta|q_X^*)- R] \} \notag \\
    & \ge \min_{q_X: R(\Delta|q_X)\geq R} \{ D(q_X\|P) -\rho[R(\Delta|q_X)- R] \}\notag\\
    & \ge \min_{q_X \in \mathcal{P(Y)} } \{ D(q_X\|P) -\rho[R(\Delta|q_X)- R] \}\notag\\
    & \stackrel{\rm (b)}
      = \rho R + \min_{q_X\in \mathcal{P(Y)}} \Big\{ D(q_X\|P) -\rho \sup_{\nu\ge 0} 
      \Big[ -\nu \Delta - \max_{p_Y} \sum_x q_X(x) \log p_Y(y) {\rm e}^{-\nu d(x,y)}
    \Big] \Big\} \notag\\
    & = \rho R + \inf_{\nu\ge 0} \min_{q_X} \max_{p_Y} \Big[ 
    \rho\nu \Delta + D(q_X\|P)    
       + \rho\sum_x q_X(x) \log \sum_{y} p_Y(y) {\rm e}^{-\nu d(x,y)}  
    \Big] \notag\\
    & \stackrel{\rm (c)}
      = \rho R + \inf_{\nu\ge 0} \max_{p_Y} \min_{q_X} \Big[ 
      \rho\nu \Delta + D(q_X\|P)    
       + \rho\sum_x q_X(x) \log \sum_{y} p_Y(y) {\rm e}^{-\nu d(x,y)}  
    \Big] \notag\\
    & \stackrel{\rm (d)}
      = \rho R + \inf_{\nu\ge 0} \bigg[ \rho\nu \Delta      
       + \max_{p_Y}-\log \sum_{x} P(x) \Big\{\sum_y p_Y(y) {\rm e}^{-\nu d(x,y)} \Big\}^{-\rho}  
    \bigg]. \label{eq.proof.lemma1} 
\end{align}
Step (a) holds because $q^*_X$ satisfies $R(\Delta|q_X) \ge R$. 
In Step (b), Eq.~(\ref{parametric_Rate_distortion}) is substituted.
Step (c) follows from the minimax theorem. It holds 
because $D(q_X\|P)$ is a convex function of $q_X$ and
$ \sum_x q_X(x) \log p_Y(y) {\rm e}^{-\nu d(x,y)}$ is linear in $q_X$ and
concave in $p_Y$. 
Step (d) holds because we have
\begin{align}
&    D(q_X\|P) + \rho\sum_x q_X(x) \log \sum_{y} p_Y(y) {\rm e}^{-\nu d(x,y)} \notag\\
&= \sum_x q_X(x) \log \frac{q_X(x)}{ P(x) \{ \sum_{y} p_Y(y) {\rm e}^{-\nu d(x,y)} \}^{-\rho} } \notag\\
&=\sum_x q_X(x) \log \frac{q_X(x)}{ \frac{1}{K} P(x) \{ \sum_{y} p_Y(y) {\rm e}^{-\nu d(x,y)} \}^{-\rho} }
-\log K\notag \\
&\stackrel{\rm (e)}
\ge -\log K,
\end{align}
where $K = \sum_{x\in \mathcal{X} } P(x) \{ \sum_{y} p_Y(y) {\rm e}^{-\nu d(x,y)} \}^{-\rho} $.
In Step (e), equality holds when $q_X(x) = \frac{1}{K} P(x) \{ \sum_{y} p_Y(y) \cdot {\rm e}^{-\nu d(x,y)} \}^{-\rho}$. 
Because Eq.~(\ref{eq.proof.lemma1}) holds any $\rho\ge 0$, we have
\begin{align}
    & E_{\rm M}(R|\Delta, P) \notag\\
    & \ge \sup_{\rho\ge 0} 
    \bigg\{
\rho R + \inf_{\nu\ge 0} \bigg[ \rho\nu \Delta      
       + \max_{p_Y} - \log \sum_{x} P(x) \Big\{\sum_y p_Y(y) {\rm e}^{-\nu d(x,y)} \Big\}^{-\rho}  
    \bigg]
    \bigg\}\notag\\
 & = E_{\rm B}(R|\Delta, P_X).    
\end{align}
This completes the proof.\hfill$\IEEEQED$

\label{Section:Proof}

\subsection{Proof of Lemma~\ref{Lemma_G_nu}}
If $\mu=0$, we have
\begin{align}
G^{(0,\nu)}(p_Y|P) 
&= \max_{q_X} 
-\sum_x q_X(x) \log \sum_y p_Y(y) {\rm e}^{-\nu d(x,y)} \notag\\
&= - \log \min_x \sum_y p_Y(y) {\rm e}^{-\nu d(x,y)}.
\end{align}
The maximum is attained by $q_X(x) = 1$ for
$x = \arg\min_x \sum_y p_Y(y) {\rm e}^{-\nu d(x,y)}$.
If $\mu>0$, we have
\begin{align}
& G^{(\mu,\nu)}(p_Y|P) \notag\\ 
&=
-\mu\min_{q_X} 
\left[
\sum_x q_X(x) \log \frac{q_X(x)}{ P(x) \left[ \sum_y p_Y(y) \mathrm{e}^{-\nu d(x,y) } \right]^{-1/\mu} }
\right] \notag \\
&= - \mu \min_{q_X} D(q_X\|q_X^*) + \mu \log K \notag\\
&= \mu \log K, \notag
\end{align}
where 
$q_X^*(x) = \frac{1}{K} P(x) \left[ \sum_y p_Y(y) \mathrm{e}^{-\nu d(x,y) } \right]^{-1/\mu}$
and 
$K= \sum_x P(x) \left\{ \sum_{y} q_Y (y) \mathrm{e}^{-\nu d(x,y)} \right\}^{-1/\mu}$.
This completes the proof.\hfill\IEEEQED

\subsection{Proof of Lemma~\ref{lemma3}}
Before describing the proof of Lemma~\ref{lemma3}, 
we show that the function $G^{(\nu)}(E, p_Y|P)$ satisfies the following property:
\begin{property} \label{property3}
For fixed $\nu\geq 0$, $p_Y$, and $P$, $G^{(\nu)}(E, p_Y | P)$ is a monotone non-decreasing and 
concave function of $E\geq 0$.
\end{property}

\

{\it Proof of Property~\ref{property3}}:
Monotonicity is obvious from the definition. Let us prove the concavity. 
Choose $E_0, E_1\geq 0$ arbitrarily. Set $E_\alpha = \alpha E_1 + (1-\alpha) E_0$
for $\alpha \in [0,1]$. Let the optimal distribution that attains 
$G^{(\nu)}(p_Y,  E_0 | P)$ and $G^{(\nu)}(p_Y,  E_1 | P)$
be $q_X^0$ and $q_X^1$. 
Then we have $D(q_X^i\|P)\leq E_i$ for $i=0,1$. 
By the convexity of the KL divergence,
we have
$ D(\alpha q_X^1 + (1-\alpha) q_X^0 \| P )
\leq
\alpha D(q_X^1  \| P ) +
(1-\alpha) D(q_X^0 \| P ) \leq
\alpha E_0 + (1-\alpha) E_1 = E_{\alpha}.
$
Therefore we have
\begin{align}
    & G^{(\mu)} (p_Y| E_\alpha, P) \notag\\
    & = \sup_{
    \genfrac{}{}{0pt}{}{q_X \in \mathcal{P(X)}:}{ D(q_X\|P) \leq E_\alpha } 
    }
    \Big\{
     -\sum_x q_X(x) \log \sum_y p_Y(y) {\rm e}^{-\nu d(x,y) }
    \Big\} \notag\\
    &\geq 
    \left. -\sum_x q_X(x) \log \sum_y p_Y(y) {\rm e}^{-\nu d(x,y) } \right|_{q_X=  \alpha q_X^1 + (1-\alpha) q_X^0}\notag\\
&= \alpha G^{(\nu)}(p_Y, E_1|P) + (1-\alpha) G^{(\nu)}(p_Y, E_0|P) .
\end{align}
This completes the proof. \hfill$\IEEEQED$


\

\textit{Proof of Lemma~\ref{lemma3}}: 
We prove that (i) for any $\mu\geq 0$ 
$ G^{(\nu)}(E, p_Y|P) \le G^{(\mu,\nu)}(p_Y|P ) + \mu E$ holds 
and (ii) there exists a $\mu\ge 0$ such that 
$ G^{(\nu)}(E, p_Y|P) \ge G^{(\mu,\nu)}(p_Y|P ) + \mu E$ holds.

Part (i): For any $\mu\geq 0$, we have
\begin{align}
    & G^{(\nu)} ( p_Y, E | P ) \notag\\
    & = \sup_{
    \genfrac{}{}{0pt}{}{q_X \in \mathcal{P(X)}:}{ D(q_X\|P) \leq E } 
    }
    \Big\{
     -\sum_x q_X(x) \log \sum_y p_Y(y) {\rm e}^{-\nu d(x,y) }
    \Big\} \notag\\
    & 
    \stackrel{\rm (a)}
    \leq    
    \sup_{
    \genfrac{}{}{0pt}{}{q_X \in \mathcal{P(X)}:}{ D(q_X\|P) \leq E } 
    }
    \Big\{
     -\sum_x q_X(x) \log \sum_y p_Y(y) {\rm e}^{-\nu d(x,y) } 
     + \mu ( E-D(q_X\|P) )
    \Big\} 
    \notag\\
    & 
    \stackrel{\rm (b)}
    \leq    
    \sup_{
        q_X \in \mathcal{P(X)}:
    }
    \Big\{
     -\sum_x q_X(x) \log \sum_y p_Y(y) {\rm e}^{-\nu d(x,y) } + \mu (E-D(q_X\|P))
    \Big\} \notag \\
    &= \mu E + G^{(\mu,\nu)} (p_X|P).
\end{align} 
Step (a) holds because adding a positive term $\mu (E-D(q_X||P))$ to the objective function 
does not decrease the supremum.
Step (b) holds because removing the restriction for the domain of the variable $q_X$ does not decrease the supremum.
Thus Part (i) is proved.

Part (ii): 
From Property~\ref{property3}, for a fixed $E\geq 0$, there exist a $\mu \geq 0$
such that for any $E'$ we have
\begin{align}
    G^{(\nu)}(p_Y, E'|P) \leq G^{(\nu)}(p_Y, E|P) + \mu (E'-E). \label{eq.36}
\end{align}
Fix this $\mu=\mu(E)$ and let $q_X'$ be a probability distribution that attains the right-hand side of (\ref{Gmunup_YP}) for this choice of $\mu$.
Put $E' = D(q_X'||P)$ for this $q_X'$ and then we have 
\begin{align}
    & G^{(\nu)}(p_Y, E|P)  \notag\\
    &
    \stackrel{\rm (a)}
    \geq G^{(\nu)}(p_Y, E'|P) - \mu ( D(q_X'||P) - E ) \notag\\
    &
    \stackrel{\rm (b)}
    \geq 
    -\sum_x q_X'(x) \log \sum_y p_Y(y) {\rm e}^{-\nu d(x,y) }  + \mu E - \mu D(q_X'\|P) \notag\\
    & 
    \stackrel{\rm (c)}
    = \mu E + G^{(\mu, \nu)} (p_Y|P). 
\end{align}
Step (a) follows form (\ref{eq.36}) and the choice of $E'$.
Step (b) holds because $q_X'$ satisfies $D(q_X'||P) \le E'$
and therefore is a feasible variable for $G^{(\nu)}(p_Y, E'|P)$. 
Step (c) holds because of the choice of $q_X'$. 
This proves Part (ii), which completes the proof. \hfill$\IEEEQED$

\fi

\subsection{Proof of Lemma~\ref{lemma:Rtilde}}
\label{Section:Proof:lemma:Rtilde}
Let $\check q_X$ be a distribution that attains $R_{\rm M}(E|\Delta, P)$ and
let $\check E = D(\check q_X\|P)$.
Then we have
\begin{align}
    & \max_{E'\le E} \widetilde R(E' |\Delta,P) \notag\\
    & \stackrel{\rm (a)} \ge 
    \widetilde R(\check E |\Delta,P) \notag \\
    & = 
        \max_{ 
        \genfrac{}{}{0pt}{}{q_X \in \mathcal{P(X)}:}{D(q_X\|P) = \check E}
        } 
        R(\Delta | q_X) \notag \\
    & \stackrel{\rm (b)} \ge
    R(\Delta|\check q_X) = R_{\rm M}(E|\Delta,P).
    \label{eqn.lemma.R_tilde.leq}
\end{align}
Step (a) follows from that $\check E$ is smaller than or equal to $E$ 
because of the definition of $\check E$.
Step (b) holds because $\check q_X$ satisfies $D(\check q_X\|P) = \check E$.

On the other hand, let $\widehat E'$ and $\hat q_X$ be
an $E'$ and $q_X$ that attains
$
\max_{E'\le E} 
\max_{ 
    \genfrac{}{}{0pt}{}{q_X \in \mathcal{P(X)}:}{D(q_X\|P) = E' }
    } 
R(\Delta | q_X)        
$. Then we have
\begin{align}
    & R_{\rm M}(E|\Delta, P) \notag\\ 
    & =
    \max_{ 
        \genfrac{}{}{0pt}{}{q_X \in \mathcal{P(X)}:}{D(q_X\|P) \le E}
    } 
    R(\Delta | q_X) \notag \\
    &
    \stackrel{\rm (a)} 
    \ge
    \max_{ 
        \genfrac{}{}{0pt}{}{q_X \in \mathcal{P(X)}:}{D(q_X\|P) \le \hat E'}
    } 
    R(\Delta | q_X) \notag\\
    & 
    \stackrel{\rm (b)}\ge 
    R(\Delta | \hat q_X)
    = \max_{E' \le E} \widetilde R(E'|\Delta, P).
    \label{eqn.lemma.R_tilde.geq}
\end{align}
Step (a) follows from that $\hat E'\le E$ because of the definition.
Step (b) follows from that $\hat q_X$ satisfies
$D(\hat q_X\|P) = \widehat E'$.

From (\ref{eqn.lemma.R_tilde.leq}) and (\ref{eqn.lemma.R_tilde.geq}), we have (\ref{eqn.lemma.R_tilde}),
completing the proof. \hfill\IEEEQED

\subsection{Proof of Lemma~\ref{lemmaB}}
To prove Lemma~\ref{lemmaB} we use the following lemma. 
\begin{lemma}[Palaiyanur and Sahai\cite{Palaiyanur2008}]
\label{Lemma:continuity_rate_distortion}
Suppose that the distortion measure $d(x,y)$ satisfies the condition mentioned in Footnote~\ref{condition_distortion_measure}.
Let $\tilde d = \min_{(x,y): d(x,y) >0} d(x,y)$. 
Then 
for any $p,q\in \mathcal{P(X)}$ such that $\lVert p-q \rVert_1 \le \tilde{d}/(4 d_{\rm max} )$
and $\Delta \ge 0$, we have
\begin{align}
    |R(\Delta| p) - R(\Delta|q)| \le -\frac{7 d_{\rm max} }{\tilde d} \lVert p-q \rVert_1 \log \frac{\lVert p-q \rVert_1 }
    {\lvert \mathcal{X}\rvert \lvert\mathcal{Y}\rvert}.
\end{align}
\end{lemma}
See~\cite{Palaiyanur2008} for the proof.

\textit{Proof of Lemma~\ref{lemmaB}:}
By Lemma~\ref{Lemma:continuity_rate_distortion}, 
there exists a $\xi'(\epsilon)>0$ such that for any $p_X, q_X \in \mathcal{P(X)}$ satisfying 
$\lVert p_X -q_X  \rVert_1 \le \xi'(\epsilon)$, 
\begin{align}
|R(\Delta|p_X) - R(\Delta|q_X ) | \le \epsilon \label{Uniform_continuity_Rate-distortion_Function}
\end{align}
holds. 
Fix this $\xi'(\epsilon)$ and put
$\xi(\epsilon) = \xi'(\epsilon)/ \sqrt{|\mathcal{X}|}$.
Then, 
due to the Cauchy-Schwartz inequality,
for any $p_X, q_X\in \mathcal{P(X)}$ satisfying 
$\lVert p_X - q_X \rVert_2 \le \xi(\epsilon)$, 
we have 
$\lVert p_X - q_X \rVert_1 \leq 
\sqrt{|\mathcal{X}|} \, 
\lVert p_X - q_X \rVert_2 
\le
\xi'(\epsilon)$.
Hence, from (\ref{Uniform_continuity_Rate-distortion_Function}), 
we have 
\begin{align}
    R(\Delta| p_X ) - \epsilon 
    \le R(\Delta|q_X ) \le  
    R(\Delta| p_X ) + \epsilon.
    \label{eq.72}
\end{align}
for any $p_X$, $q_X \in \mathcal{P(X)}$ with $\lVert p_X - q_X \rVert_2 \le \delta(\epsilon)$.
Taking the minimum and the maximum values of $R( \Delta | q_X )$ w.r.t. $q_X$,
we obtain (\ref{lemmaBmax}) and (\ref{lemmaBmin}).
This completes the proof.\hfill\IEEEQED

\subsection{Proof of Lemma~\ref{lemmaA}}
To prove Lemma~\ref{lemmaA}, we use the following lemma. 
\begin{lemma}
    \label{L_1bound_on_Divergence}
    Fix a $P\in \mathcal{P(X)}$.
    For any $p,q\in \mathcal{P(X)}$ such that $p(x)=q(x) =0$ if $P(x)=0$. 
    If $\lVert p-q \rVert_1 \leq \frac12$,
    we have
    \begin{align}
        \lvert D(p \| P) - D(q \| P) \rvert \notag
        & \le - \lVert p-q \rVert_1 \log \frac{\lVert p-q \rVert_1}{\lvert \mathcal{X} \rvert } \notag\\
        &\quad - \lVert p-q \rVert_1 \log \min_{
        \genfrac{}{}{0pt}{}{x\in \mathcal{X}: }{P(x)>0}
        }
        P(x)
    \end{align}    
\end{lemma}

\textit{Proof:}
Since $D(p\|P) = - H(p) - {\rm E}_{p} [ \log P(X) ]$ holds,
$\lvert D(p \| P) - D(q \| P) \rvert$ is upper bounded by
$ |H(p) -H(q)| + | {\rm E}_{p} [ \log P(X) ] 
- {\rm E}_{q} [ \log P(X) ] 
|$. 
We use the $L_1$ bound on entropy~\cite[Theorem 17.3.3]{CoverTEXT}, i.e.,
$
|H(p) - H(q)|\leq - \lVert p-q \rVert_1 
    \log \frac{\lVert p-q \rVert_1}{|\mathcal{X}|} 
$ if $\lVert p-q \rVert_1\leq 1/2$.
Let $r(x) = |p(x) - q(x) |$. 
The second term is upper bounded as follows.
\begin{align}
&| {\rm E}_{p} [ \log P(X) ] 
- {\rm E}_{q} [ \log P(X) ] 
| \notag\\
&=
    \left|
    \sum_{x\in \mathcal{X} } 
    ( p(x)-q(x) ) \log P(x) 
    \right| \notag \\
    &
    \leq
    -\sum_{x\in \mathcal{X}}
    r(x) \log P(x) \notag \\
    &
    =
    - \lVert p-q \rVert_1 
    \sum_{x\in \mathcal{X}}
    \frac{r(x)}{ \lVert p-q \rVert_1 }
    \log P(x)    \notag \\
    &
    \le - \lVert p-q \rVert_1 
    \log \min_{
    \genfrac{}{}{0pt}{}{x\in \mathcal{X}: }{P(x)>0}
    } P(x).
\end{align}
The proof is completed by adding the upper bounds of the two terms.\hfill\IEEEQED

\textit{Proof of Lemma~\ref{lemmaA}:}
Fix $E\in [0, E^*]$ and $\xi>0$ arbitrarily. 
Assume, for the sake of contradiction, that for any 
positive integer $n$, there exists a $p_X^{(n)}$ 
satisfying
\begin{align}
    | D(p_X^{(n)} \| P ) - E | \le \delta_n
    \label{Bounds_D_2}
\end{align}
with $\delta_n = 1/n$ such that for any
$q_X\in \mathcal{P(X)}$ satisfying $D(q_X\|P)=E$, 
we have
\begin{align}
    \lVert p_X^{(n)} - q_X \rVert_2 > \xi.  
\end{align}
Since the space $\mathcal{P(X)}$ is a closed bounded subset of $\mathbb{R}^{|\mathcal{X}|}$,
it follows from the Bolzano-Weierstrass theorem that the sequence $\{p_X^{(n)}\}$ has 
a subsequence converging to an element in $\mathcal{P(X)}$. That is, there exists
$q_X^* \in \mathcal{P(X)}$ and $k_n$, $n=1,2,\ldots$ such that
\begin{align}
    \lVert p_X^{(k_n)} - p_X^* \rVert_2 \to 0, n \to \infty
    \label{convergence_L2norm2}
\end{align}
For any $|\mathcal{X}|$-dimensional vector $r$, the Cauchy-Schwarz inequality gives 
$\lVert r \rVert_1 \leq \sqrt{\lvert \mathcal{X} \rvert }\,  \lVert r \rVert_2$. 
Thus, (\ref{convergence_L2norm2}) implies
\begin{align}
    \| p_X^{(k_n)} - p_X^* \|_1 \to 0, \quad n\to \infty.
    \label{convergence_L1norm}
\end{align}
Note that $p_X$ for which $D(p_X\|P)$ is finite satisfies $p_X(x) = 0$ if $P(x)=0$. 
Then, it follows from Lemma~\ref{L_1bound_on_Divergence} and (\ref{convergence_L1norm}) that
\begin{align}
   | D(p_X^{(k_n)} \|P ) - D(p_X^{*} \|P ) |\to 0, \quad n\to \infty. 
\end{align}
From (\ref{Bounds_D_2}) and the continuity of $D(p_X\|P)$ in $p_X\in \mathcal{P(X)}$, 
we have 
\begin{align}
    \lim_{n\to \infty} D(p_X^{(k_n)} \|P ) = D(p_X^{*} \|P ) = E.    
\end{align}
Then, by choosing $q_X = p_X^*$, we have 
$D(q_X\|P) = E$ and 
\begin{align}
    \lVert p_X^{(n)} - q_X \rVert_2 = 
    \lVert p_X^{(n)} - p_X^* \rVert_2 \to 0, \quad n\to \infty.
\end{align}
This is a contradiction. Hence the first assumption was wrong and
the statement of the lemma holds. This completes the proof.
\hfill\IEEEQED

\end{document}
\ifISIT
\else 

\newpage 
\appendices


\section{Proof of Lemmas~\ref{lemma.1} and \ref{lemma:Rtilde}}
\label{appendix_proof}
This section gives the proofs of Lemmas~\ref{lemma.1} and \ref{lemma:Rtilde}.

\end{document}


\begin{lemma}[(to be deleted)] \label{lemma:choice_of_delta}
Fix an $\hat E \geq 0$ and an $\epsilon>0$. 
Put $\hat R = R_{\rm M}( \hat E| \Delta, P)$. 
Then we can choose a $\delta>0$ such that the two sets
\begin{align}
    &\{q_X \in \mathcal{P(X)}: R(\Delta|q_X) \ge \hat R + \epsilon\}, \label{setQ}\\
    &\{q_X \in \mathcal{P(X)}: D(q_X \| P ) \le \hat E + \delta\}\label{setP}
\end{align}
are disjoint.
\end{lemma}

\textit{Proof:} 
Assume, for the sake of contradiction, that, for any positive integer $n$,
there exists a $ q_X^{(n)} \in P(\mathcal{X})$ such that 
\begin{align}
    R(\Delta | q_X^{(n)}) & \ge \hat R + \epsilon, \label{assumption} \\
    D(q_X^{(n)} \|P) & \le \hat E + \delta_n. \label{assumption2}
\end{align}
with $\delta_n=1/n$. 
Since $\hat R$ is the maximum of $R(\Delta|q_X)$ 
over $\{q_X: D(q_X\|P) \le \hat E \}$, 
(\ref{assumption}) implies $D(q_X^{(n)} \|P) > \hat E$.
This, combined with (\ref{assumption2}), gives
\begin{align}
    \hat E < D(q_X^{(n)} \|P) \le \hat E + {\textstyle \frac1n} . \label{Bounds_D}
\end{align}

Since the space $\mathcal{P(X)}$ is a closed bounded subset of $\mathbb{R}^{|\mathcal{X}|}$, 
it follows from the Bolzano–Weierstrass theorem that the sequence 
$\{ q_X^{(n)} \}$ has a subsequence converging to an element in $\mathcal{P(X)}.$
That is, there exists $q_X^*\in \mathcal{P(X)}$ and $k_n$, $n=1,2,...$ such that
\begin{align}
    \| q_X^{(k_n)} - q_X^* \|_2 \to 0, \quad n\to \infty.
    \label{convergence_L2norm}
\end{align}
For any $|\mathcal{X}|$-dimensional vector $r$, the Cauchy-Schwarz inequality gives 
$\lVert r \rVert_1 \leq \sqrt{\lvert \mathcal{X} \rvert }\,  \lVert r \rVert_2$. 
Thus, (\ref{convergence_L2norm}) implies
\begin{align}
    \| q_X^{(k_n)} - q_X^* \|_1 \to 0, \quad n\to \infty.
    \label{convergence_L1norm}
\end{align}
Note that $q_X$ for which $D(q_X\|P)$ is finite satisfies $q_X(x) = 0$ if $P(x)=0$. 
Then, it follows from Lemma~\ref{L_1bound_on_Divergence} and (\ref{convergence_L1norm}) that
\begin{align}
   | D(q_X^{(k_n)} \|P ) - D(q_X^{*} \|P ) |\to 0, \quad n\to \infty. 
\end{align}
From (\ref{Bounds_D}) and the continuity of $D(q_X\|P)$ in $q_X\in \mathcal{P(X)}$, 
we have 
\begin{align}
    \lim_{n\to \infty} D(q_X^{(k_n)} \|P ) = D(q_X^{*} \|P ) = \hat E.    
\end{align}
Then, from Lemma~\ref{Lemma:continuity_rate_distortion} and (\ref{assumption}), we have
\begin{align}
    \lim_{n\to \infty} R(\Delta | q_X^{(k_n)}) 
    & = R(\Delta | q_X^*) \ge \hat R + \epsilon.
\end{align}
However, by the definition of $R_{\rm M}(\hat E|\Delta, P)$, we must have
\begin{align}
    D(q_X\|P)\le \hat E \implies R(\Delta|q_X)\leq \hat R.
\end{align}
This is a contradiction. Hence the first assumption was wrong
and the statement of the lemma holds. This completes the proof.
\hfill\IEEEQED

\fi

\end{document}